\documentclass[pra,twocolumn,amsmath,amssymb,superscriptaddress,
showpacs]{revtex4}
\usepackage{graphics}
\usepackage{epsfig}
\usepackage{bbm}
\usepackage[latin1]{inputenc} 
\usepackage{subfigure}

\newcommand{\be}{\begin{equation}}
\newcommand{\ee}{\end{equation}}
\newcommand{\eea}{\end{eqnarray}}
\newcommand{\bea}{\begin{eqn array}}

\newcommand{\va}[1]{\ensuremath{(\Delta#1)^2}}

\newcommand{\ex}[1]{\ensuremath{\left\langle{#1}\right\rangle}}
\newcommand{\exs}[1]{\ensuremath{\langle{#1}\rangle}}

\newcommand{\qed}{\ensuremath{\hfill \blacksquare}}

\newcommand{\ketbra}[1]{\ensuremath{| #1 \rangle \langle #1 |}}

\newcommand{\ket}[1]{\ensuremath{|#1\rangle}}

\newcommand{\bra}[1]{\ensuremath{\langle#1|}}

\newcommand{\kommentar}[1]{}
\newcommand{\trace}{{\rm Tr}}

\renewcommand{\vr}{\ensuremath{\rho}}
\newcommand{\singlet}{\Psi^-}
\def\be{\begin{equation}}
\def\ee{\end{equation}}
\def\eea{\end{eqnarray}}
\def\bea{\begin{eqnarray}}

\newcommand{\coll}{J}
\newcommand{\collB}{j}
\newcommand{\nmatrix}{Q}

\newcommand{\observation}[1]{{\bf Observation #1.}---}

\newcommand{\aver}[1]{\langle #1 \rangle}

\newcommand{\id}{\openone}

\newcommand{\SSIs}{spin-squeezing inequalities }

\newcommand{\spsubs}{{\rm os}}

\begin{document}
\title{
Spin squeezing and entanglement for arbitrary spin}
\date{\today}
\begin{abstract}
A complete set of generalized spin-squeezing inequalities is derived for an ensemble of particles with an arbitrary spin. Our conditions are formulated with the first and second moments of the collective angular momentum coordinates.  A method for mapping the spin-squeezing inequalities for spin-$\frac{1}{2}$ particles to entanglement conditions for spin-$j$ particles is also presented. We 
apply our mapping to obtain a generalization of the original spin-squeezing inequality to higher spins. 
We show that, for large particle numbers, a spin-squeezing parameter for entanglement detection based on one of our inequalities is
 strictly stronger than
the original spin-squeezing parameter defined in [A. S\o rensen {\it et al.}, Nature {\bf 409}, 63 (2001)].
We present a coordinate system independent form of our inequalities that contains, besides the correlation and covariance tensors of the collective angular momentum operators, the nematic tensor appearing in the theory of spin nematics.
Finally, we discuss how to measure the quantities appearing in our inequalities in experiments.
\end{abstract}

\author{Giuseppe Vitagliano}
\affiliation{Department of Theoretical Physics, University of the Basque Country UPV/EHU,
P.O. Box 644, E-48080 Bilbao, Spain}

\author{Iagoba Apellaniz}
\affiliation{Department of Theoretical Physics, University of the Basque Country UPV/EHU,
P.O. Box 644, E-48080 Bilbao, Spain}

\author{I\~nigo L. Egusquiza}
\affiliation{Department of Theoretical Physics, University of the Basque Country UPV/EHU,
P.O. Box 644, E-48080 Bilbao, Spain}

\author{G\'eza T\'oth}
\email{toth@alumni.nd.edu}
\homepage[\\URL: ]{http://www.gtoth.eu}
\affiliation{Department of Theoretical Physics, University of the Basque Country  UPV/EHU,
P.O. Box 644, E-48080 Bilbao, Spain}
\affiliation{IKERBASQUE, Basque Foundation for Science, E-48011 Bilbao, Spain}
\affiliation{Wigner Research Centre for Physics, Hungarian
Academy of Sciences, P.O. Box 49, H-1525 Budapest, Hungary}



\maketitle



\section{Introduction}

One of the most rapidly developing areas in quantum physics is creating larger and 
larger entangled quantum systems with photons, trapped ions, and cold neutral atoms 
\cite{review,KS07,WK09, CG12, LZ07,HH05, MS11,JK01,revatom,MG03,IA12,MB12}. 
Entangled states can be used for metrology in order to obtain a sensitivity higher than the 
shot-noise limit \cite{SD01,PS09,fisher_kentanglement} and can also be used as a 
resource for certain quantum information processing tasks \cite{RB03,G96,G97,crypto}. Moreover, 
experiments realizing macroscopic quantum effects 
might give answers to fundamental 
questions in quantum physics \cite{D89,FD12}.

Spin squeezing is one of the most successful approaches for creating large-scale quantum entanglement
\cite{K93,W94,HS99,VR01,SD01,SM99,WS01,KB98,HM04,HP06,EM05,EM08,F08,Achip,BEC,BEC2,BEC3}.
It is used in systems of very many particles in which only collective quantities can be measured. For 
 an ensemble of $N$ particles with a spin $j,$ the most relevant collective quantities 
 are the collective spin operators defined as 
\begin{equation}
J_l:= \sum_{n=1}^N j_l^{(n)} 
\end{equation}
for $l=x,y,z,$ where  $j_l^{(n)}$ are the components of the angular momentum operator 
for the $n^{\rm th}$ spin.

Spin-squeezed states are typically almost fully polarized states for which the angular momentum
variance is small in a direction orthogonal to the mean spin \cite{K93}. They can be used
to achieve a high accuracy in certain very general metrological tasks
\cite{PS09,fisher_kentanglement}.
On the other hand, in spin-$\frac{1}{2}$ systems spin squeezing is closely connected to multipartite entanglement. A ubiquitous criterion for detecting the entanglement of spin-squeezed states is \cite{SD01}
\begin{eqnarray}
\xi_{\rm s}^2:=N \frac{\va{J_x}}{\exs{J_y}^2+\exs{J_z}^2}\ge 1.
\label{motherofallspinsqueezinginequalities}
\end{eqnarray}
Any fully separable state of $N$ qubits, that is, a state that can be written as \cite{W89}
\begin{equation}\label{sep}
\varrho=\sum_k p_k \varrho_k^{(1)}\otimes \varrho_k^{(2)}\otimes ... \varrho_k^{(N)}, \;\;\;\sum_k p_k=1, \;\;\;p_k>0,\\
\end{equation}
satisfies Eq.~(\ref{motherofallspinsqueezinginequalities}). Any state violating Eq.~(\ref{motherofallspinsqueezinginequalities}) is not fully separable and is therefore entangled. 

Apart from the original inequality Eq.~(\ref{motherofallspinsqueezinginequalities}), several other generalized spin-squeezing entanglement conditions have been presented \cite{RevNori,GT04,WVB05,KC05,KC06,GT06,KK09,CP11,SM01,HP11,DC02,MZ02,MK08,Relquadspin,DuanPRL,MGD12}
and even the complete set of such criteria for multi-qubit systems has been found in Ref.~\cite{TK07}.
While most of the conditions are for a fixed particle number, conditions for 
the case of nonzero particle number variance have also been derived \cite{HP10,HP12}.

So far most of the attention has been focused on ensembles of 
spin-$\frac{1}{2}$ particles.
The literature on systems of particles with $j>\frac{1}{2}$ has been
limited to a small number of conditions, specialized
to certain sets of quantum states or particles with a low spin
\cite{SM01,GT04,WVB05,MK08,HP11,DC02,MZ02}. The reason is that known methods for detecting entanglement for spin-$\frac{1}{2}$ particles by spin-squeezing cannot straightforwardly
be generalized to higher spins. For example, for $j > \frac{1}{2},$
Eq.~(\ref{motherofallspinsqueezinginequalities})
can also be violated without entanglement between the spin-$j$ particles, as we will discuss later \cite{F08}. 

In spite of the  difficulties in deriving entanglement conditions for particles with a higher spin, they are very much needed
in quantum experiments nowadays. As most of such experiments are done with atoms with $j>\frac{1}{2},$
such conditions can make the complexity of experiments much smaller:  The artificially created spin-$\frac{1}{2}$ subsystems must be manipulated by lasers, while the physical spin-$j$ particles can directly be manipulated by magnetic fields.
Moreover higher spin systems could make it possible to perform quantum information processing tasks different from the ones possible with spin-$\frac{1}{2}$ particles
or to create different kind of entangled states \cite{TM10,UH12,KTqutrit,LS11,HG12,RP01,SU12}.

In this paper, we will start from the complete set presented for  spin-$\frac{1}{2}$ particles in Ref.~\cite{TK07}.
All spin-squeezing entanglement criteria of this set are based on the first and second moments of collective angular momentum coordinates. It has been possible to obtain a full set of tight inequalities by analytical means only due to certain advantageous properties of the spin-$\frac{1}{2}$ case. 
For the case of particles with $j > \frac{1}{2},$ the inequalities presented in the literature are either based on numerical optimization \cite{SM01} or are analytical but not tight \cite{MK08}. The reason for this is that for $j > \frac{1}{2},$ the second moments of the collective observables are not only connected to the two-body correlations, as in the spin-$\frac{1}{2}$ case, but also to the local second moments.

In order to solve this problem, we define modified second moments and the corresponding variances as follows
\begin{align}\label{tildequant}
\aver{\tilde{J}_{l}^{2}} &:= \aver{J_{l}^{2}} - \aver{\sum_{n} (j_{l}^{(n)})^{2}} 
=\sum_{n \neq m} \aver{j_l^{(n)} j_l^{(m)}} , \notag \\
\left( \tilde{\Delta} J_l \right)^{2} &:= \aver{\tilde{J}_l^{2}} - \aver{J_l}^{2},
\end{align}
where $l=x,y,z.$ 
The modified quantities do not contain anymore the local second moments.
We will show that by using the first moments and
the modified second moments of the collective operators,
it is possible to write down tight entanglement conditions
analytically also for the $j > \frac{1}{2}$ case \cite{Note_tilde}.
We will also discuss that the local second moments are related to
single-particle spin squeezing (see Sec.~\ref{singlepart_spinsq}).

\begin{figure}
\centerline{ \epsfxsize3.2in
\epsffile{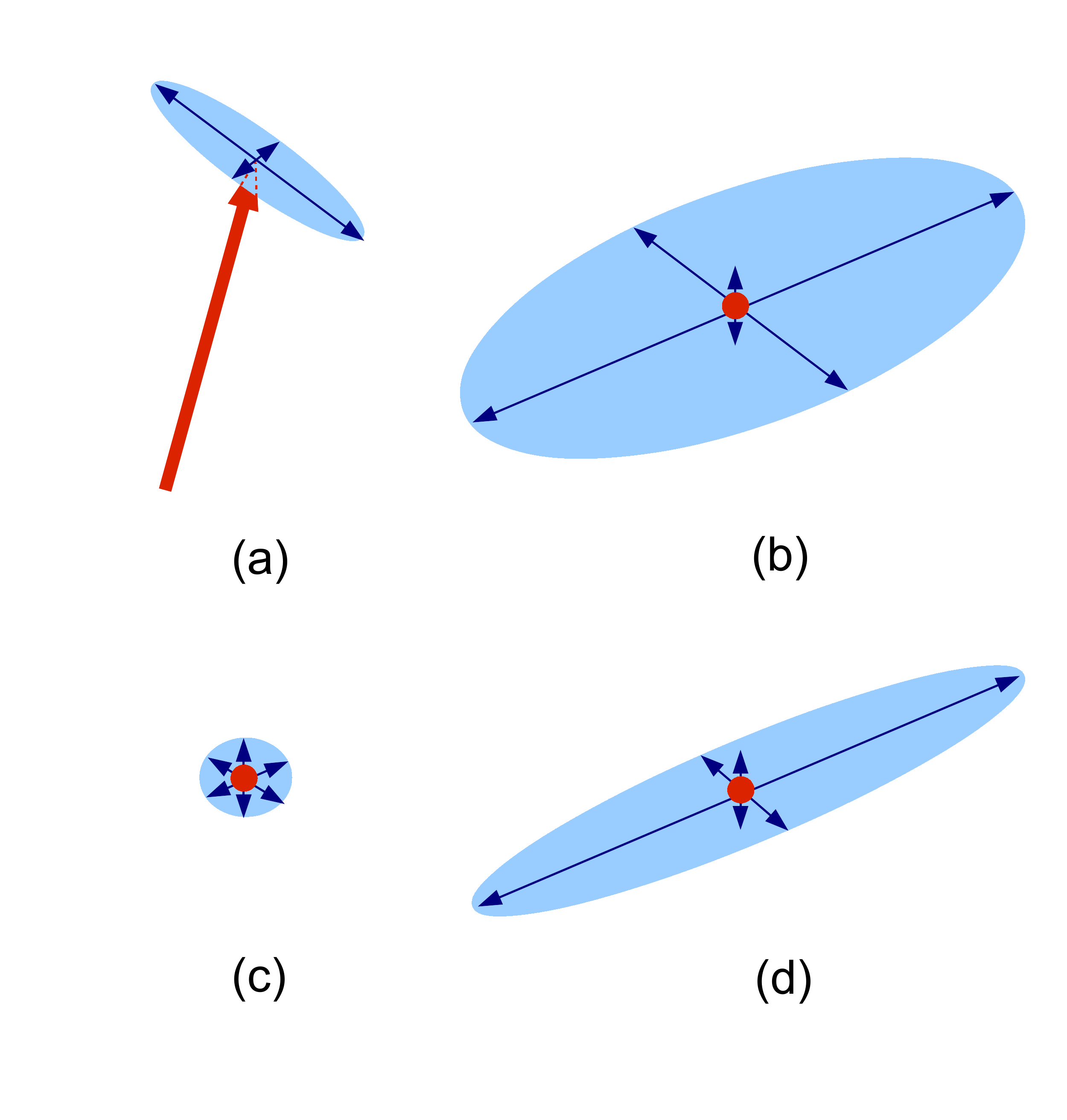}
} 
\caption{(Color online) Different types of spin-squeezed states. (a) {\it Almost fully polarized spin-squeezed states} detected by $\xi_{{\rm s},j}^2,$ given in Eq.~(\ref{transosp1}) 
and also by the new parameter, $\xi_{\spsubs}^2,$ defined in Eq.~(\ref{transosp11}).
(b) States close to {\it symmetric Dicke states} with $\ex{J_z}=0$ with a small variance for one of the angular momentum components and large variances in the two orthogonal directions. Such states can be detected by  $\xi_{\spsubs}^2$ but  are not detected by $\xi_{{\rm s},j}^2.$
(c) States close to {\it many-body singlets} with a small variance for all the three angular momentum components.
Such states are detected by the criterion~(\ref{isoin}).
(d) {\it Planar squeezed states} with a small variance for two of the angular momentum components and a large variance in the orthogonal direction. Such states are detected by the criterion~(\ref{twovar}).
} \label{spisqfig} 
\end{figure}

The main results of our paper are as follows. 

(i) We will find the complete set of conditions for  the  $j > \frac{1}{2}$ case, which 
we will call {\it optimal spin-squeezing inequalities for spin-$j$ particles}.
They are a complete set since, for large $N,$  they detect all entangled states that can be 
detected knowing  only the first moments and the modified second moments.
For instance, they can be used to verify the entanglement of
 singlet states, symmetric Dicke states and planar squeezed 
states \cite{HP11}.

(ii) We also present a generalization of the original spin
squeezing parameter $\xi_{{\rm s}}^2$ defined in  Eq.~(\ref{motherofallspinsqueezinginequalities})  that can be used for entanglement detection even for particles with $j>\frac{1}{2}$
\begin{equation}\label{transosp1}
\xi_{{\rm s},j}^2:=N \dfrac{( \tilde{\Delta} J_{x})^{2}+Nj^2}{\aver{J_{y}}^{2}+\aver{J_{z}}^{2}}.
\end{equation}
If $\xi_{{\rm s},j}^2<1$ then the state is entangled.
 For spin-$\frac{1}{2}$ particles, the definitions of Eqs.~(\ref{motherofallspinsqueezinginequalities}) and
 (\ref{transosp1}) are the same. 

(iii) Finally, we will show that, in the large particle number limit, 
the entanglement condition based on the following entanglement parameter 
\begin{equation}\label{transosp11}
\xi_{\spsubs}^2:= (N-1) \dfrac{( \tilde{\Delta} J_{x})^2+Nj^2}{\aver{\tilde{J}_{y}^2}+\aver{\tilde{J}_{z}^2}} 
\end{equation}
is strictly stronger than the condition based on $\xi_{{\rm s},j}^2.$ 
Note that $\xi_{\spsubs}^2$ is defined only for $\aver{\tilde{J}_{y}^2}+\aver{\tilde{J}_{z}^2}>0.$ In this way $\xi_{\spsubs}^2$ will always be non-negative. In Eq.~(\ref{transosp11}), the subscript ``$\spsubs$" refers to the optimal spin-squeezing inequalities since
we obtain $\xi_{\spsubs}^2$, essentially, by dividing the left-hand side of one of the inequalities by the right-hand side.
For clarity, we give Eq.~(\ref{transosp11}) explicitly for the $j=\frac{1}{2}$ case 
\begin{equation}\label{transosp11bbb}
\xi_{\spsubs}^2=(N-1)\dfrac{( {\Delta} J_{x})^2}{\aver{{J}_{y}^2}+\aver{{J}_{z}^2}-\frac{N}{2}} .
\end{equation}
If $\xi_{\spsubs}^2<1$ then the state is entangled.
The parameter (\ref{transosp1}) is appropriate only for spin-squeezed states with a large total spin depicted in Fig.~\ref{spisqfig}(a), while
the parameter (\ref{transosp11}) detects also states that have zero total spin, as shown in Fig.~\ref{spisqfig}(b).
Moreover, we will also show that for large particle numbers, if $\xi_{{\rm s},j}^2<1$ then we also have
\begin{equation}\label{ineqso1}
\xi_{\spsubs}^2 < \xi_{{\rm s},j}^2.
\end{equation}
Thus, $\xi_{\spsubs}^2$ is a better indicator of entanglement than $\xi_{{\rm s},j}^2.$

The paper is organized as follows. In Ref.~\cite{PRL}, we have already presented a generalization of the complete set of 
spin-squeezing inequalities 
valid for systems of spin-$j$ particles with $j>\frac{1}{2}.$
In this paper, we extend the results of Ref.~\cite{PRL} in several directions.
In Sec.~\ref{complset}, we present the optimal \SSIs for spin-$j$ particles and discuss some of their fundamental properties.
In Sec.~\ref{statesvio}, we study states that violate the inequalities maximally. 
In Sec.~\ref{comparis}, we show a method for mapping existing entanglement conditions
for spin-$\frac{1}{2}$ particles to analogous conditions for spin-$j$ particles with $j>\frac{1}{2}.$ 
Using the mapping, we derive the spin-squeezing parameter $\xi_{{\rm s},j}^2$.
In Sec.~\ref{sec_a_stronger}, we present the spin-squeezing parameter $\xi_{\spsubs}^2$ and examine its
properties.
In Sec.~\ref{further}, we consider various issues concerning the
efficient application of our spin-squeezing inequalities.

\section{Complete set of spin-squeezing inequalities for spin-$j$ particles.}\label{complset}

In this section, we present our spin-squeezing inequalities for particles with an arbitrary spin  $j$
and we also examine the connection of these inequalities to the entanglement of the reduced two-particle state,
and to the criterion based on the
positivity of the partial transpose.

\subsection{The optimal spin-squeezing inequalities for qudits }\label{optisecA}

\observation{1}The following generalized spin-squeezing inequalities are 
valid for separable states given by Eq.~(\ref{sep}) for an ensemble of 
spin-$j$ particles even with  $j>\frac{1}{2}$ 
\begin{subequations}\label{spinjssi}
\begin{align}
\aver{J_{x}^{2}}+\aver{J_{y}^{2}}+\aver{J_{z}^{2}} &\leq Nj(Nj+1) , & \label{symmsatin} \\
(\Delta J_{x})^{2}+(\Delta J_{y})^{2}+(\Delta J_{z})^{2} &\geq Nj , & \label{isoin}  \\
\aver{\tilde{J}_{l}^{2}} + \aver{\tilde{J}_{m}^{2}}  -N(N-1)j^{2} &\leq (N-1)  (\tilde{\Delta} J_{k})^{2}, &
 \label{betosp}  \\
(N-1) \left[ (\tilde{\Delta} J_{k})^{2} + (\tilde{\Delta} J_{l})^{2} \right] &\geq  \aver{\tilde{J}_{m}^{2}}-N(N-1)j^{2} . &  \label{twovar} 
\end{align}
\end{subequations}
Here $k,l,m$ may take all the possible permutations of $x,y,z.$ If a quantum state violates one of the inequalities (\ref{spinjssi}), then it is entangled.

{\it Proof.} We will prove that for separable states the following inequality holds
\begin{equation}\label{ssij}
(N-1) \sum_{l \in I} (\tilde{\Delta} J_l)^{2} - \sum_{l \notin I}\aver{\tilde{J}_l^{2}} \geq -N(N-1)j^{2} ,
\end{equation}
where $I$ is a subset of indices including the two extremal cases $I=\emptyset$ and $I=\{x,y,z\}$.
We consider first pure product states of the form
$\ket{\Phi} = \otimes_n \ket{\phi_n}.$
For such states, the modified variances and the modified second moments can be obtained as
\begin{eqnarray}
\va{\tilde{\coll}_l}_{\Phi} &=&-\sum_{n}\exs{\collB_l^{(n)}}^2,\nonumber\\
\ex{\tilde{\coll}_l^2}_{\Phi} &=&\exs{J_l}^2-\sum_{n}\exs{\collB_l^{(n)}}^2=\sum_{n\ne m} \exs{\collB_l^{(n)}}\exs{\collB_l^{(m)}}. \nonumber\\\label{subs}
\end{eqnarray}
Substituting Eq.~(\ref{subs}) into the left-hand side of Eq.~(\ref{ssij}), we obtain
\begin{eqnarray}
&&-\sum_{n}(N-1)\sum_{l\in I}\exs{\collB_{l}^{(n)}}^{2}
-\sum_{l\notin I}\left(\exs{\coll_{l}}^{2}-
\sum_{n}\exs{\collB_{l}^{(n)}}^{2}\right)\nonumber\\
&&\;\;\;\;\;\;\;\;\;\;\;\ge
-\sum_{n}(N-1)\sum_{l=x,y,z}\exs{\collB_{l}^{(n)}}^{2}\ge-N(N-1)j^2.\nonumber\\
\label{opt_gen}
\end{eqnarray}
The two inequalites in Eq.~(\ref{opt_gen}) follow from the inequality  \cite{TK07}
\begin{equation}
\exs{\coll_{l}}^{2}\le
N\sum_{n}\exs{\collB_{l}^{(n)}}^{2},
\end{equation}
and from the well-known bound for an angular momentum component 
$\langle j_l\rangle\le j.$
Hence we proved that Eq.~(\ref{ssij}) is valid for pure product states.
Due to the left-hand side of Eq.~(\ref{ssij}) being concave in the state, it is also valid 
for separable states.

From Eq.~(\ref{ssij}) we can obtain all inequalities of Eq.~(\ref{spinjssi}a)-(\ref{spinjssi}d), knowing that
\begin{equation}\label{J2xyzJtilde2xyz}
 \exs{J_x^2}+\exs{J_y^2}+\exs{J_z^2}=\exs{\tilde{J}_x^2}+\exs{\tilde{J}_y^2}+\exs{\tilde{J}_z^2}+Nj(j+1),
\end{equation}
which is a consequence of  the identity \cite{angular}
\begin{eqnarray}
 j_{x}^{2}+ j_{y}^{2}+ j_{z}^{2}&=&j(j+1) \id. \label{jxyz2}
\label{algspin}
\end{eqnarray} 
Hence, we proved that Eq.~(\ref{spinjssi}) is valid for separable states.
\qed\\

In order to evaluate Eq.~(\ref{spinjssi}), six operator expectation values are needed. 
These are the vector of the expectation values of the three collective angular momentum components
\begin{equation}\label{Jvec}
\vec J:=(\aver{J_{x}},\aver{J_{y}},\aver{J_{y}}),
\end{equation}
and the vector of the modified second moments 
\begin{equation}
\vec{\tilde{K}}:=(\aver{{\tilde{J}_{x}^2}},\aver{{\tilde{J}_{y}^2}},\aver{\tilde{J}_{y}^2}). \label{tildeK}
\end{equation}
For the spin-$\frac{1}{2}$ case, the modified second moments can be obtained from the true second moments since $\aver{{\tilde{J}_{x}^2}}=\aver{{{J}_{x}^2}}-\frac{N}{4}.$
For spin-$j$ particles with $j>\frac{1}{2},$ the elements of $\vec{\tilde{K}}$ typically cannot be measured directly. Instead, we measure the true second moments
\begin{equation}\label{Kvec}
\vec{K}:=(\aver{{{J}_{x}^2}},\aver{{{J}_{y}^2}},\aver{{J}_{y}^2})
\end{equation}
and the sum of the squares of the local second moments
\begin{equation}\label{Mvec}
\vec{M}:=\left(\aver{\sum_n (j_x^{(n)})^2},\aver{\sum_n (j_y^{(n)})^2},\aver{\sum_n (j_z^{(n)})^2}\right). 
\end{equation}
Then, $\vec{\tilde{K}}$ can be obtained as the difference between the true second moments and the sum of local second moments as 
\begin{equation}
\vec{\tilde{K}}=\vec{K}-\vec{M}.
\end{equation}
In Sec.~\ref{addcom}, we discuss how to measure $\vec{\tilde{K}}$ based on the measurement 
of $\vec{K}$ and $\vec{M}.$

\begin{figure}
\centerline{ \epsfxsize3.3in \epsffile{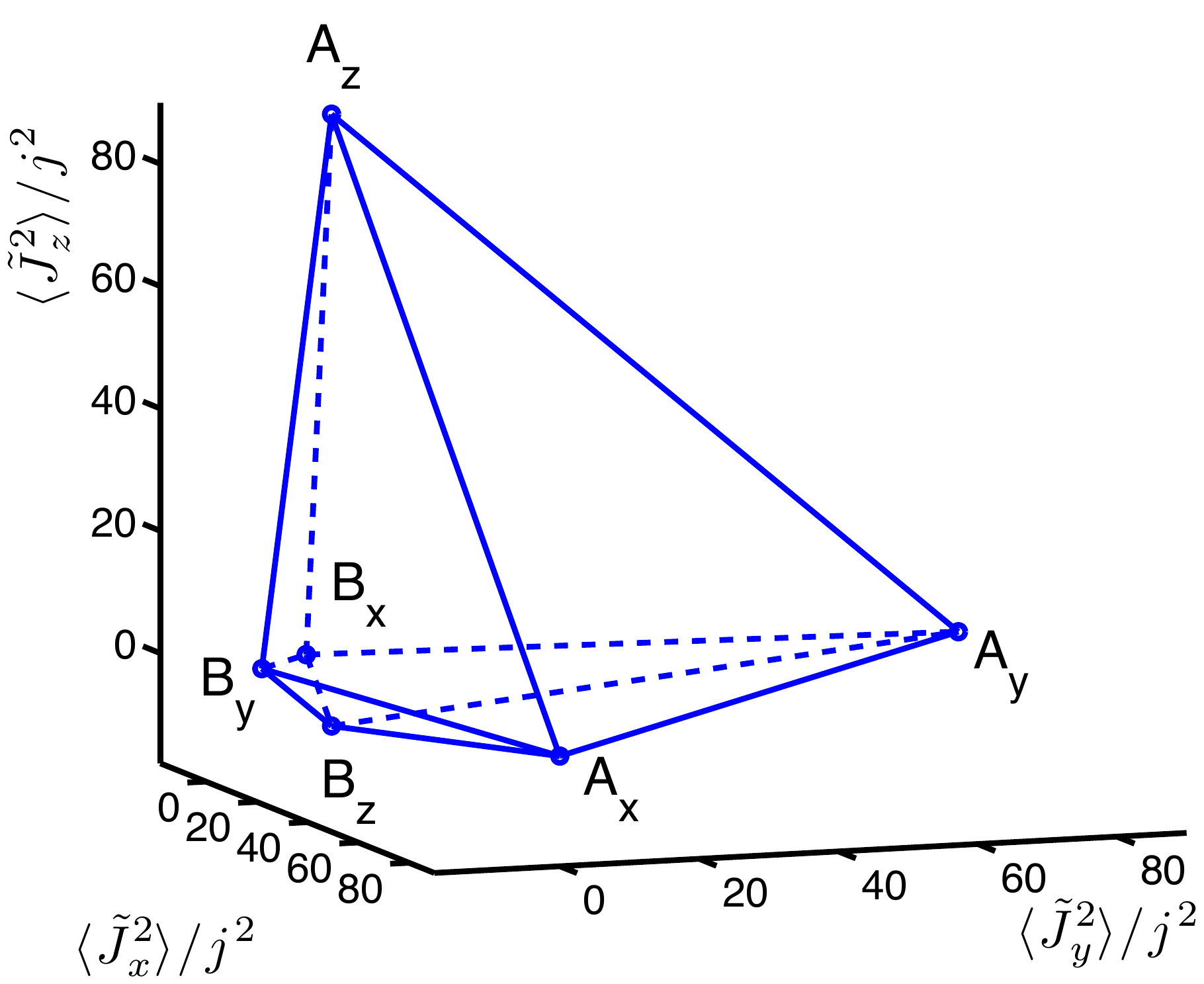}}
\centerline{\bf \large(a)}
\vskip0.5cm
\centerline{ \epsfxsize3.3in \epsffile{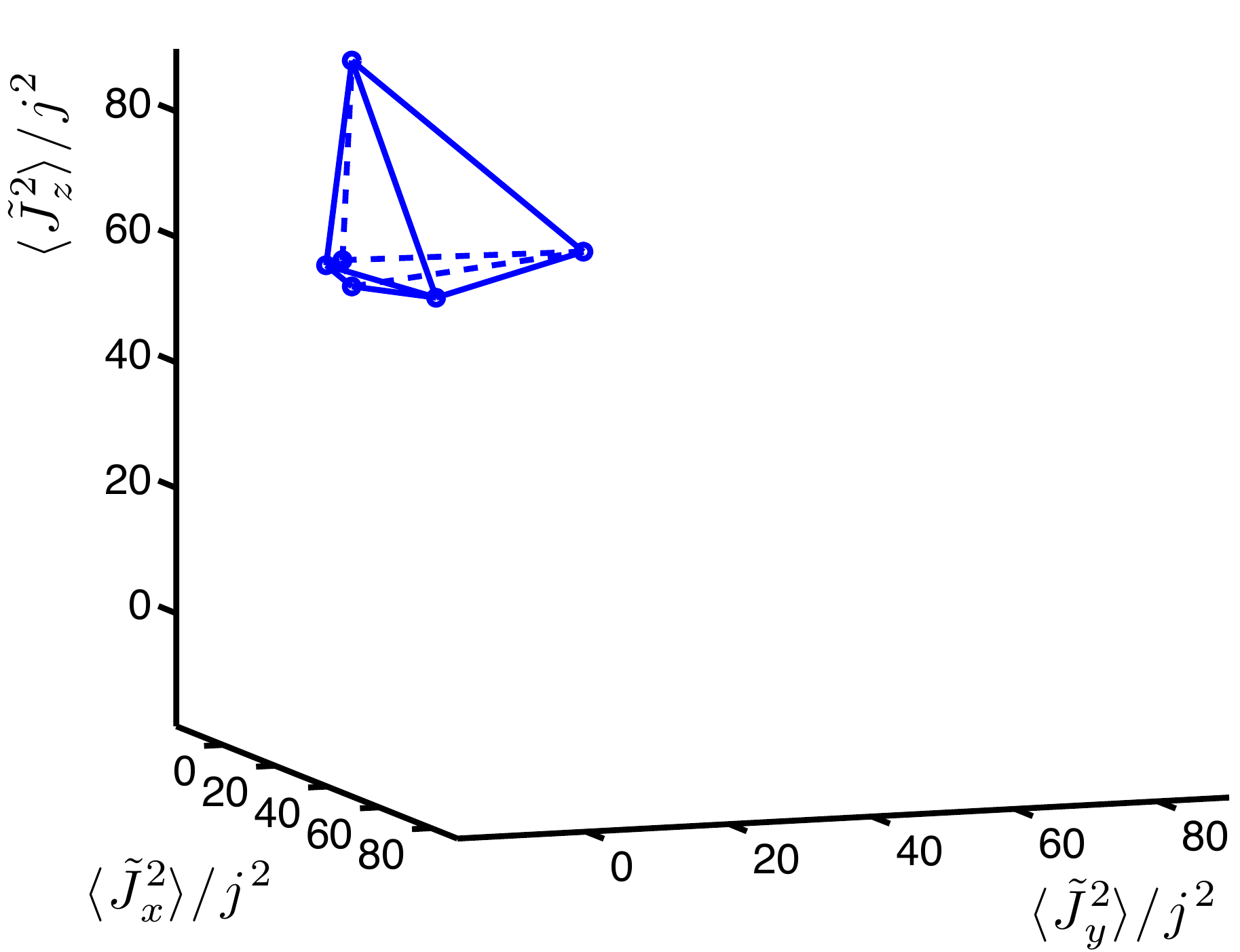}}
\centerline{\bf \large(b)}
\caption{(Color online) (a) The polytope of separable states corresponding to
Eqs.~(\ref{spinjssi})  for $N=10$ spin-$j$ particles and for $\vec{J}=0.$ The completely mixed state defined in Eq.~(\ref{cm}) corresponds to the origin of the coordinate axis, i.e., the point $(0,0,0)$ and it is inside the polytope. (b)
The same polytope for $\vec{J}=(0,0,8)j.$ Note that this polytope is
a subset of the polytope in (a). For the coordinates of the points $A_l$  and $ B_l$  see Eq.~(\ref{AB}).  }
\label{J2xyz}
\end{figure}

For any value of the mean spin $\vec J,$
Eq.~(\ref{spinjssi})  defines a polytope in the $(\aver{\tilde{J}_x^2},\aver{\tilde{J}_y^2},\aver{\tilde{J}_z^2})$-space.
The polytope
is depicted in Figs.~\ref{J2xyz}(a) and \ref{J2xyz}(b) for different values for
$\vec{J}.$ It is completely characterized by its extremal points.
Direct calculation shows that the coordinates of the extreme points
in the $(\exs{\tilde{J}_{x}^2},\exs{\tilde{J}_{y}^2},\exs{\tilde{J}_{z}^2})$-space are
\begin{align}
A_x &:=\left[ N(N-1)j^2-\kappa(\exs{J_y}^2+\exs{J_z}^2),
\kappa\exs{J_y}^2,\kappa\exs{J_z}^2
\right], \nonumber
\\
B_x&:=\left[ \exs{J_x}^2+\frac{\exs{J_y}^2+\exs{J_z}^2}{N}-Nj^2,
\kappa \exs{J_y}^2,\kappa\exs{J_z}^2
\right], \nonumber\\
\label{AB}
\end{align}
where $\kappa:=\frac{N-1}{N}.$ The points $A_{y/z}$ and $B_{y/z}$ can be
obtained in an analogous way. Note that the coordinates of the
points $A_l$ and $B_l$ depend nonlinearly on $\exs{J_l}.$

Let us see briefly the connection between the inequalities and the facets of the polytope.
The inequality with three second moments,  
Eq.~(\ref{symmsatin}), corresponds to the facet $A_x-A_y-A_z$ in Fig.~\ref{J2xyz}(a). 
The inequality with three variances, Eq.~(\ref{isoin}), corresponds to the facet $B_x-B_y-B_z.$ The inequality with one variance,
Eq.~(\ref{betosp}) corresponds to the facets $B_x-A_y-A_z,$
$B_y-A_z-A_x,$  and $B_z-A_x-A_y.$  
The inequality with two variances, Eq.~(\ref{twovar}), corresponds to the facets $B_x-B_y-A_z,$ 
$B_y-B_z-A_x$ and $B_z-B_x-A_y.$ 

\subsection{Completeness of Eq.~\eqref{spinjssi}} \label{optisecC}

In this section, we will show that, in the large $N$ limit, all points inside the polytope correspond to
separable states. This implies that the criteria of Observation
1 are complete, that is, if the inequalities are not violated then
it is not possible
to prove the presence of entanglement 
based only on the first and the modified second moments.
In other words, it is not possible to find criteria
detecting more entangled states based on these moments. 
To prove this, first we can observe that
if some quantum states satisfy Eq.~\eqref{spinjssi} then their mixture also
satisfies it. Thus, it is enough to
investigate the states corresponding to the extremal points of the polytope.
We will give a straightforward generalization of 
the proof for the spin-$\frac{1}{2}$ case presented in Ref.~\cite{TK07}.

\observation{2}(i) For any value of $\vec{J}$ there are
separable states corresponding to $A_k$ for $k\in\{x,y,z\}.$\\
(ii) Let us define $J:=Nj,$ \be
c_x:=\sqrt{1-\tfrac{\exs{J_y}^2+\exs{J_z}^2}{J^2}},\ee and
$p:=\frac{1}{2}[1+\tfrac{\exs{J_x}}{J c_x}].$ If $Np$ is an integer then
there exists also a separable state corresponding to $B_x.$ Similar
statements hold for $B_y$ and $B_z.$ Note that this condition is
always fulfilled, if $\vec{J}=0$ and $N$ is even.
\\
(iii) There are always separable states corresponding to points
$B_k'$ such that their distance from $B_k$ is smaller than
$j^2.$ In the limit $N\rightarrow \infty$ for a fixed
normalized angular momentum $\tfrac{\vec{J}}{N},$ 
the points $B_k$ and the $B_k'$ cannot be distinguished by 
measurement, for that a precision $j^2$ or better would be needed
when measuring $\exs{\tilde{J}_x^2},$ which is unrealistic.
Hence in the macroscopic limit the characterization is complete. 

{\it Proof.} A separable state corresponding to $A_x$ is
\begin{equation}
{\vr}_{A_x}:=p(\ketbra{\psi_+})^{\otimes N}+
(1-p)(\ketbra{\psi_-})^{\otimes N}. \label{Ax}
\end{equation}
Here $\ket{\psi_{+/-}}$ are the single-particle states with $(\exs{j_x},\exs{j_y},\exs{j_z})=j
(\pm c_x,\frac{\exs{J_y}}{J},\frac{\exs{J_z}}{J})$.

If $M:=Np$ is an integer, we can
also define the state corresponding to the point $B_x$ as \be
\ket{\phi_{B_x}}:=\ket{\psi_+}^{\otimes M}\otimes
\ket{\psi_-}^{\otimes (N-M)}. \label{Bx} \ee Since there is a
separable state for each extreme point of the polytope, for any
internal point a corresponding separable state can be obtained by
mixing the states corresponding to the extreme points. 

If $M$ is not an integer, we can approximate $B_x$ by taking $m:=
M-\varepsilon$ as the largest integer smaller than $M,$ defining the state
\begin{eqnarray}\vr'&:=&(1-\varepsilon) (\ketbra{\psi_+})^{\otimes m}
\otimes (\ketbra{\psi_-})^{\otimes(N-m)}  \nonumber \\&+&\varepsilon
(\ketbra{\psi_+})^{\otimes (m+1)} \otimes (\ketbra{\psi_-})^{\otimes
(N-m-1)}.\nonumber\\
\end{eqnarray}
It has the same
coordinates as $B_x,$ except for the value of $\exs{\tilde{J}_x^2},$ where
the difference is $ 4j^2 c_x^2 \varepsilon(1- \varepsilon) \le j^2.$  \qed

The extremal states that correspond to the vertices of the polytope defined by the optimal \SSIs
are, in a certain sense, generalizations of the coherent spin states defined as \cite{RevNori,Bcoh}
\begin{eqnarray}
\ket{\Psi_{\rm CSS}}=\ket{\Psi}^{\otimes N}, \label{css}
\end{eqnarray}
where $\ket{\Psi}$ is a state with maximal $\aver{j_x}^2+\aver{j_y}^2+\aver{j_z}^2.$
All states of the form (\ref{css}) saturate all the inequalities, as 
can be seen by direct substitution into Eq.~(\ref{spinjssi}).
Further extremal states can be obtained as tensor products or mixtures of  coherent spin states.
Note that they exist for all the possible values of the mean spin $\vec J,$
while spin coherent states Eq.~(\ref{css}) were fully polarized. 

\subsection{Relation of Eq.~\eqref{spinjssi} to two-particle entanglement}\label{ave2part}

Since the optimal spin-squeezing inequalities~(\ref{spinjssi}) contain only first moments and modified second moments of the angular momentum components,
they can be reformulated with the average two-body correlations.
For that, we define the average two-particle density matrix as
\begin{equation}
\label{avertwopar}
\rho_{{\rm av2}}:= \tfrac 1 {N(N-1)} \sum_{m\neq n} \rho_{mn} ,
\end{equation}
where $\rho_{mn}$ is the two-particle reduced density matrix for the $m^{\rm th}$ and $n^{\rm th}$ particles.

Next, we formulate our entanglement conditions with the density matrix $\rho_{{\rm av2}}.$

\observation{3}The optimal spin-squeezing inequalities Eq.~(\ref{spinjssi}) for arbitrary spin can be given in terms of the
average two-body density matrix as 
\begin{gather}\label{tq}
N \sum_{l\in I} \left( 
\aver{j_{l} \otimes j_{l}}_{{\rm av2}} - \aver{j_{l}\otimes\openone}_{{\rm av2}}^{2}  \right) \geq \Sigma- j^2,
\end{gather}
where we have defined the expression $\Sigma$ as the sum of all the two-particle correlations of the local spin operators
\begin{equation}
\Sigma := \sum_{l=x,y,z} \aver{j_{l} \otimes j_{l}}_{{\rm av2}}.
\end{equation}
The right-hand side of Eq.~(\ref{tq}) is nonpositive.  For the $j=\frac{1}{2}$ case, the right-hand side of Eq.~(\ref{tq}) is zero for all symmetric states,
while for $j>\frac{1}{2}$ it is zero only for some symmetric states. 

{\it Proof.} Equation~(\ref{ssij}) can be transformed into
\begin{equation}\label{symmexp1}
N \sum_{l \in I} (\tilde{\Delta} J_l)^{2} + \sum_{l\in I}\aver{{J}_l}^{2} \geq \sum_{l}\aver{\tilde{J}_l^{2}}-N(N-1)j^2.
\end{equation}

Next, let us see how Eq.~(\ref{symmexp1}) behaves for symmetric states.
We know from angular momentum theory that Eq.~(\ref{symmsatin}) of the optimal \SSIs
is saturated only when the state is symmetric. For the $j=\frac{1}{2}$ case, all symmetric states saturate Eq.~(\ref{symmsatin}), 
while for $j>\frac{1}{2}$ only some of the symmetric states saturate it. 
Based on these and Eq.~(\ref{J2xyzJtilde2xyz}),
we know that, for spin-$\frac{1}{2}$ particles in a symmetric state the right-hand side of 
Eq.~(\ref{symmexp1}) is zero. On the other hand,
for spin-$j$ particles with $j>\frac{1}{2}$ in a symmetric state, the right-hand side can also be negative.
 
 Let us now turn to the reformulation of Eq.~(\ref{symmexp1})  in terms of the two-body reduced density matrix.
 The modified second moments and variances can be expressed with the average two-particle density matrix as
\begin{eqnarray}
\aver{\tilde{J}^{2}_l} &=& \sum_{m\neq n} \aver{j_{l}^{(n)} j_{l}^{(m)} } = 
N(N-1) \aver{j_{l} \otimes j_{l}}_{{\rm av2}} , \nonumber \\
(\tilde{\Delta}J_l)^{2}  
&=& - N^{2} \aver{j_{l}\otimes\openone}_{{\rm av2}}^{2} + 
N(N-1) \aver{j_{l}\otimes j_{l}}_{{\rm av2}} .\nonumber \\\label{twopart}
\end{eqnarray}
Substituting Eq.~(\ref{twopart}) into Eq.~(\ref{symmexp1}), we obtain Eq.~(\ref{tq}).
As in the case of Eq.~(\ref{symmexp1}), the right-hand side of Eq.~(\ref{tq}) is zero for symmetric states  of spin-$\frac{1}{2}$ particles.
$\qed$

Not that, as in the spin-$\frac{1}{2}$ case, there are states detected as entangled that have a separable two-particle density matrix \cite{TK07}. 
Such states are, for example, permutationally invariant states with certain symmetries for which the reduced single-particle density matrix is completely mixed.
For large $N,$ due to permutational invariance and the symmetries mentioned above, the two-particle density matrices are  very close to the a completely mixed matrix as well and hence they are separable. Still, some of such states can be detected as entangled by the optimal spin-squeezing inequalities. Examples of such states are the permutationally invariant singlet states discussed later in Sec.~\ref{subsec_ineq_three_variances}.

\subsection{Relation of Eq.~\eqref{spinjssi} to the criterion based on the positivity of the partial transpose}

 Our inequalities are entanglement conditions.
 Thus, it is important to compare them to the most useful entanglement condition known so far, the condition based on the positivity of the partial transpose (PPT) \cite{PPT}.

In Ref.~\cite{TK07}, it has been shown for the spin-$\frac{1}{2}$ case that the optimal \SSIs can detect the thermal states of some spin models
 that have a positive partial transpose for all bi-partitions of the system. Such states are extreme forms of bound entangled states: they are non-distillable even if the qubits of the two partitions are allowed to unite with each other. We found that for the $j>\frac{1}{2}$ case,
 the inequality~(\ref{isoin}) also detects such bound entangled states in the thermal states of spin models.
An example of such a state for $j=1$ and $N=3$ is
\begin{equation}
\varrho_{\rm BES}\propto e^{-\frac{J_x^2+J_y^2+J_z^2}{T}}.\label{BES}
\end{equation}
The state (\ref{BES}) is detected by our criterion below the temperature bound $T_{\rm s}\approx 3.66$ while
it is detected by the PPT criterion below the bound $T_{\rm PPT}\approx 3.57.$
  
Finally, we will consider the special case of symmetric states. In this case, the PPT condition 
applied to the reduced two-body density matrix detects all states detected by the spin-squeezing inequalities.
 
\observation{4}The PPT criterion for the average two-particle density matrix defined in Eq.~(\ref{avertwopar})  
detects all symmetric entangled states that the optimal \SSIs detect for $j>\frac{1}{2}.$ The two conditions are equivalent for symmetric states of particles with  $j=\frac{1}{2}.$ 

{\it Proof.}  We will connect the violation of Eq.~(\ref{tq}) to the violation 
of the PPT criterion by the reduced two-particle density matrix $\varrho_{\rm av2}.$ 
If a quantum state is symmetric, its
reduced state $\varrho_{\rm av2}$ is also symmetric. For such states, 
the PPT condition is equivalent to \cite{TG09}
\begin{equation}\label{PPT}
\aver{A \otimes A}_{\rm av2} - \aver{A\otimes \id}_{\rm av2}^{2} \geq 0
\end{equation}
holding for all Hermitian operators $A.$ Based on Observation 3, it can be seen by straightforward comparison of Eqs.~(\ref{tq}) and (\ref{PPT}) that, for  $j=\frac{1}{2},$  Eq.~(\ref{tq})  holds for all possible choices of $I$ and for all possible choices of 
coordinate axes, i.e., all possible $j_l,$ if and only if
Eq.~(\ref{PPT})  holds  for all Hermitian operators $A.$ For  $j>\frac{1}{2}$ there is no equivalence between the two statements. Only from the latter follows the former. \qed

\section{States that violate the optimal spin-squeezing inequalities for spin $j$}\label{statesvio}

In this section we will study, what kind of states violate maximally our spin-squeezing inequalities.
We will also examine, how much noise can be mixed with these states such that they are still
detected as entangled by our inequalities.

\subsection{The inequality with three second moments, Eq.~(\ref{symmsatin})}
\label{subsec_ineq_three_second_moments}

The first two equations of Eqs.~(\ref{spinjssi}) are invariant under the exchange of coordinate axes $x,y,$ and $z$. As a consequence of basic angular momentum theory, Eq.~(\ref{symmsatin}), the inequality with three second moments is valid for all quantum states, thus it cannot be violated.
As discussed in the proof of Observation 3, for the $j=\frac{1}{2}$ case, all symmetric states saturate Eq.~(\ref{symmsatin}), 
while for $j>\frac{1}{2}$ only some of the symmetric states saturate it. 
In both cases, states of the form (\ref{css}) are a subset of the saturating states.

\subsection{The inequality with three variances, Eq.~(\ref{isoin})}
\label{subsec_ineq_three_variances}

The states maximally violating Eq.~(\ref{isoin}) are the many-body singlet states.
The characteristic values of the collective operators for many-body singlets
are shown in Table I.
States violating Eq.~(\ref{isoin})  have a small variance for all the components of the angular
momentum as shown in Fig.~\ref{spisqfig}(c).

Let us see now some examples of many-body singlets states.
For $j=\frac{1}{2},$ a pure singlet state can be constructed, 
for example, as a tensor product of two-particle singlets of the form 
\bea
\ket{\singlet}=\tfrac{1}{\sqrt{2}}\left(\ket{+\tfrac{1}{2},-\tfrac{1}{2}}_z-\ket{-\tfrac{1}{2},+\tfrac{1}{2}}_z\right).
\eea
Any permutation of such a state is a singlet as well. The mixture of all such permutations
is a permutationally invariant singlet defined as
\begin{equation} 
\rho_{\rm s,PI} = \tfrac{1}{N!}\sum_{k=1}^{N!} \Pi_k ( \ket{\singlet} \bra{\singlet} 
\otimes \cdots \otimes \ket{\singlet} \bra{\singlet})\Pi_k^{\dagger}, \label{singlet_pairs}
\end{equation}
where $\Pi_k$ are all the possible permutations of the qubits.
It can be shown that for even $N,$ Eq.~(\ref{singlet_pairs}) equals
 the $T=0$ thermal ground state of the Hamiltonian \cite{TM10,UH12}
\begin{equation}
H_{\rm s}=J_{x}^{2}+J_{y}^{2}+J_{z}^{2}. \label{Hxyz}
\end{equation}
For even $N$ and  $j=\frac{1}{2},$ the state $\rho_{\rm s,PI}$ is the only permutationally invariant singlet state.
For  $j=\frac{1}{2},$ all singlets are outside of the symmetric subspace.

In the case of spin-$1$ particles,  the following two-particle symmetric  state  
\begin{equation}\label{singstat}
| \phi_{\rm s1} \rangle = \tfrac 1 {\sqrt 3} \left(|1,-1\rangle - |0,0\rangle + |-1,1 \rangle \right) ,
\end{equation}
is also a singlet. 
It is very important from the point of view of experimental realizations
with Bose-Einstein condensates that for $j>\frac{1}{2}$ there are singlet states
in the symmetric subspace. 

Next, we mix the spin-$j$ singlet state with white noise and examine up to how much noise it is still violating 
Eq.~(\ref{isoin}). The noisy singlet state is the following 
\begin{equation}\label{noisysinglet}
\varrho_{{\rm s}, {\rm noisy}}(p_{\rm n})=(1-p_{{\rm n}})
\varrho_{\rm s}  +p_{{\rm n}}\varrho_{\rm cm},
\end{equation}
where $\varrho_{\rm s}$ is a singlet state maximally violating Eq.~(\ref{isoin}), and
$p_{{\rm n}}$ is the amount of noise and we defined the completely mixed state as
\begin{equation}
\rho_{{\rm cm}} = \tfrac{1}{d^N}\openone,\label{cm}
\end{equation}
where the dimension of the qudit is $d=2j+1.$
The vectors of the collective quantities $({\vec J}_{\rm cm},{\vec K}_{\rm cm},{\vec M}_{\rm cm})$ are shown in Table I for the completely mixed state.
Based on these, simple calculations show that the state (\ref{noisysinglet}) is detected as entangled by Eq.~(\ref{isoin}) if
\begin{equation}
p_{{\rm n}} < \tfrac 1 {j+1}= \tfrac{2}{d+1}.
\end{equation}
Hence, the white-noise tolerance decreases with $d$. 

\begin{table*}[htdp]
\caption{Expectation values of collective quantities appearing in the 
optimal spin-squeezing inequalities (\ref{spinjssi}) for various quantum states. 
$\vec J,$ $\vec K,$ and $\vec M$ are defined in Eqs.~(\ref{Jvec}), (\ref{Kvec}), and (\ref{Mvec}), respectively.}
\begin{tabular*}{\textwidth}{@{\extracolsep{\fill}\vspace{0.03cm}}ll}
 \hline \vspace{-0.2cm}\\
Singlet state discussed in Sec.~\ref{subsec_ineq_three_variances}& 
$\vec J_{\rm s}=(0,0,0)$ \\
 & $\vec K_{\rm s}=(0,0,0)$ \\
  & $\vec M_{\rm s}=(\frac{j(j+1)}{3}N,\frac{j(j+1)}{3}N,\frac{j(j+1)}{3}N)$  \vspace{0.2cm}\\
 \hline  \vspace{-0.2cm}\\
Completely mixed state defined in Eq.~(\ref{cm})& $\vec J_{\rm cm}=(0,0,0)$ \\
    & $\vec K_{\rm cm}=(\frac{j(j+1)}{3}N,\frac{j(j+1)}{3}N,\frac{j(j+1)}{3}N)$ \\
               & $\vec M_{\rm cm}=(\frac{j(j+1)}{3}N,\frac{j(j+1)}{3}N,\frac{j(j+1)}{3}N)$ \vspace{0.2cm}\\
\hline  \vspace{-0.2cm}\\
Symmetric Dicke state, $\vert D_{N,j}\rangle,$ discussed in Sec.~\ref{betopstat}
& $\vec J_{\rm D}=(0,0,0)$ \\
      & $\vec{ {K}}_{\rm D}=(\frac{Nj(Nj+1)}{2},\frac{Nj(Nj+1)}{2},0)$ \\
                & 
$\vec M_{\rm D}=(\frac{Nj(j+1)}{2}-\frac{N(N-1)j^2}{4jN-2},\frac{Nj(j+1)}{2}-\frac{N(N-1)j^2}{4jN-2},\frac{N(N-1)j^2}{2jN-1})$ \vspace{0.2cm}\\
\hline 
\end{tabular*}
\label{default}
\end{table*}

Finally note that for any $j$ the modified second moments of the collective angular momentum components are zero for the 
completely mixed state, i.e.,  
\begin{eqnarray}
\vec{\tilde{K}}_{\rm cm}&=&(0,0,0).
\end{eqnarray}
Thus, the completely mixed state belongs to a point at the 
origin of the coordinate system of the modified second moments for $\vec J=0.$
In contrast, in the space of true second moments the singlet state is at the origin, since for the singlet we have
$\exs{J_l^2}=\exs{J_l}=0$ for $l=x,y,z.$

Eq.~(\ref{isoin}) has been proposed to detect entanglement in optical lattices of cold atoms \cite{GT04}. A related inequality was presented for entanglement detection in condensed matter systems by susceptibility measurements \cite{WVB05}. 
Experimentally, it has been used for entanglement detection in photonic systems \cite{IA12} and in fermionic cold atoms \cite{MB12}. An ensemble of $d$-state fermions naturally fills up the energy levels of a harmonic oscillator such that all levels have $d$ fermions in a multipartite $SU(d)$ singlet state. Such a state is also a singlet, maximally violating the optimal spin-squeezing inequality with three variances, Eq.~(\ref{isoin}). Singlets can also be obtained through spin squeezing in cold atomic ensembles \cite{TM10,UH12}. 
Finally, the ground state of the system Hamiltonian for certain spinor Bose-Einstein condensates is a singlet state \cite{HG12}.

\subsection{The inequality with only one variance, Eq.~(\ref{betosp})}\label{betopstat}

Next, we will consider  the optimal spin-squeezing inequality with one variance Eq.~(\ref{betosp}).
This entanglement criterion is very useful to detect the almost fully polarized spin-squeezed states
 shown in Fig.~\ref{spisqfig}(a). It can also be used to detect 
 symmetric Dicke states with a maximal $\langle J_x^2+J_y^2+J_z^2\rangle$ and $\langle J_z\rangle=0.$
States close to such symmetric Dicke states  have a small variance for one component of the angular
momentum while they have a large variance in two orthogonal directions as shown in Fig.~\ref{spisqfig}(b).

Dicke states $\vert \lambda,\lambda_z,\alpha\rangle$ are quantum states obeying the eigenequations
\begin{eqnarray} 
(J_x^2+J_y^2+J_z^2) \vert \lambda,\lambda_z,\alpha\rangle &=& \lambda(\lambda+1) \vert \lambda,\lambda_z,\alpha\rangle,\nonumber\\
J_z \vert \lambda,\lambda_z,\alpha\rangle &=& \lambda_z \vert \lambda,\lambda_z,\alpha\rangle,
\end{eqnarray}
where $\alpha$ is a label used to distinguish the different 
eigenstates corresponding to the same eigenvalues $\lambda$ and $\lambda_z.$
In particular, we will show that Eq.~(\ref{betosp}) is very useful to detect entanglement close to the symmetric 
Dicke state
\begin{equation} \label{DNj}
\vert D_{N,j}\rangle:=\vert Nj,0\rangle,
\end{equation}
where $N$  must be even for half integer $j$'s.
In this case, the $\alpha$ label is not needed, as
the two eigenvalues determine the state uniquely. The state
state (\ref{DNj}) for $j=\frac{1}{2}$ has already been known to have intriguing entanglement properties \cite{GT06} and it is optimal for certain very general quantum metrological tasks \cite{fisher_kentanglement}.

We will now show that the state (\ref{DNj}) maximally violates Eq.~(\ref{betosp}) for 
$j=\frac{1}{2}$ and is close to violating it maximally for $j>\frac{1}{2}.$ In order to show this, we
rewrite Eq.~(\ref{betosp}) for $(k,l,m)=(z,x,y)$ as
\begin{eqnarray}
&&\aver{J_x^2+J_y^2+J_z^2}-N(\Delta J_z)^2 -\aver{J_z}^2
+N\sum_n \aver{(j_z^{(n)})^2} \nonumber\\
&&\;\;\;\;\;\;\;\;\;\;\;\;\;\;\;\;\;\;\;\;\;\;\;\;\;\;\;\;\;\;\;\;\;\;\;\;\;\;\;\;\;\;\;\;\;\leq Nj(Nj+1).
  \label{betosp2}
\end{eqnarray}
The state (\ref{DNj}) maximally violates  Eq.~(\ref{betosp}) for $j=\frac{1}{2}$ since it maximizes all terms with a positive coefficient and minimizes all terms with a negative one on the left-hand side of Eq.~(\ref{betosp2}). 
This statement is almost true also for the case $j>\frac{1}{2},$ except for the term with the local second moments which has a value 
\begin{equation}\label{jx2Dicke}
\sum_n \aver{(j_z^{(n)})^2}= \tfrac{N(N-1)j^2}{2jN-1}.
\end{equation}
The proof of Eq.~(\ref{jx2Dicke}) is given in the appendix.
Based on these, our symmetric Dicke state is detected as entangled for any $j.$

In a practical situation, it is also important to know how much additional noise is 
tolerated such that the noisy state is still detected as entangled.
Next, we look at the noise tolerance of the inequality~(\ref{betosp2}) for our case. We mix
the symmetric Dicke state (\ref{DNj}) with white noise as
\begin{equation}
\varrho_{{\rm D}, {\rm noisy}}(p_{\rm n})=(1-p_{{\rm n}})
|D_{N,j}  \rangle\langle D_{N,j} |  +p_{{\rm n}}\varrho_{\rm cm} .
\end{equation}
The expectation values  and the relevant moments of the collective angular momentum components for the 
Dicke state (\ref{DNj}) are given in Table I.
Based on these, a noisy Dicke state is detected as entangled if
\begin{equation}
p_{{\rm noise}}< \tfrac{N}{N(2j+1)-1} .
\end{equation}
For large $N,$  the bound on the noise is $\frac{1}{2j+1}.$

Entangled states close to Dicke states have been observed 
in photonic experiments with a condition similar to the optimal spin-squeezing  
inequality with one variance, Eq.~(\ref{betosp}) \cite{KS07,WK09,CG12}. 
Symmetric Dicke states can be created dynamically in Bose-Einstein condensate \cite{LS11,HG12}.
Cold trapped ions also seem to be ideal to create symmetric Dicke states, thus the use of our inequalities is expected even in these systems \cite{HH05,KC06,UF03}. 

\subsection{The inequality with two variances, Eq.~(\ref{twovar})}

As the last case let us consider the optimal spin-squeezing inequality~(\ref{twovar}). 
Typical states strongly violating Eq.~(\ref{twovar})  have a small variance for two components of the angular
momentum while having a large variance in the orthogonal direction, see Fig.~\ref{spisqfig}(d).
As we will see, for certain values for $j,$ singlet states [Fig.~\ref{spisqfig}(c)] also violate  Eq.~(\ref{twovar}). 

Now it is hard to compute the maximally
violating state, because  an independent optimization for the different terms does not seem to lead to
a state maximizing the whole expression even for $j=\frac{1}{2}.$ 
Thus, we will consider examples of important states violating the inequality and
compare it to other similar conditions.

Let us consider the multi-particle spin singlet states. Based on ${\vec J}_{\rm s},$ ${\vec K}_{\rm s},$ and 
${\vec M}_{\rm s}$ given in Table I,
 we find that the optimal spin-squeezing inequality~(\ref{twovar}) is violated whenever
\begin{equation}\label{jN}
j < \tfrac{2N-3} N .
\end{equation}
Thus, for $N\ge7,$ the singlet state is violating this inequality for $j=\frac{1}{2}, 1,$ and 
$\frac{3}{2}.$

An alternative of the entanglement condition with two variances (\ref{twovar}),
the planar squeezing entanglement condition \cite{HP11,PC12},
is of the form
\begin{equation}
\va{J_x}+\va{J_y} \ge N C_j,\label{planar}
\end{equation}
where the constant $C_j$  is $\frac{1}{4}$ for $j=\frac{1}{2}$ and $\frac{7}{16}$  for $j=1,$ respectively. For larger $j,$ the constant $C_j$ is determined numerically.  
For even $N,$ the criterion~(\ref{planar}) is maximally violated by the many-particle singlet state for any $j.$

Let us compare the entanglement condition (\ref{twovar})  
to the planar squeezing entanglement condition~(\ref{planar}).  
Using Eq.~(\ref{jxyz2}), Eq.~(\ref{twovar}) can be rewritten
for $(k,l,m)=(x,y,z)$ as
\begin{align}
& (\Delta J_{x})^{2} + (\Delta J_{y})^{2} 
\geq Nj  + \tfrac{1}{N-1}\aver{ J_{z}^{2}}-\tfrac{N}{N-1}M_z. \label{othtwo1bbb} 
\end{align}
For $j=\frac{1}{2}$ and for large $N,$ it can be seen that the right-hand side of  Eq.~(\ref{othtwo1bbb})
equals $\frac{N}{4}+\frac{1}{N-1}\exs{J_z^2}.$ A comparison with Eq.~(\ref{planar}) shows that 
our condition~(\ref{othtwo1bbb}) is strictly stronger in this case. 
For $j>\frac{1}{2},$ Eq.~(\ref{othtwo1bbb}) is not strictly stronger any more,
but still is more effective in detecting quantum states with a large $\exs{J_z^2}.$

This seems to be the advantage of our inequality compared to Eq.~(\ref{planar}):
It has information not only about the variances in the $x$ and $y$ directions, but also
about the second moment in the third direction.

\section{Spin-$\frac{1}{2}$ entanglement criteria transformed to higher spins}\label{comparis}

In this section, we present a method to map spin-$\frac{1}{2}$ entanglement criteria to criteria for higher spins.
We use it to transform the original spin-squeezing parameter Eq.~(\ref{motherofallspinsqueezinginequalities}) to a spin-squeezing parameter for higher spins. We show that two of the optimal spin-squeezing inequalities are strictly stronger than the transformed original spin-squeezing criterion. We also convert some other spin-$\frac{1}{2}$ entanglement criteria to criteria for higher spins.

\subsection{The original spin-squeezing parameter for higher spins}

Next, we present a mapping that
can transform every spin-squeezing inequality for an ensemble of spin-$\frac{1}{2}$ particles written in terms of the first and the modified second moments of the collective spin operators to an entanglement condition for spin-$j$ particles, also given in terms of the first and the modified second moments.

\observation{5}Let us consider an entanglement condition (i.e., a necessary condition for separability)
 for spin-$\frac{1}{2}$ particles of the form
\begin{equation}\label{funcj}
f(\{ \aver{J_{l}} \}, \{\aver{\tilde{J}_{l}^{2}} \} ) \geq \mathrm{const.} ,
\end{equation}
where $f$ is a six-dimensional function. Then, the inequality
obtained from Eq.~(\ref{funcj}) by the substitution
\begin{equation}\label{mapj}
\aver{J_{l}} \longrightarrow \tfrac 1 {2j} \aver{J_{l}}  , \qquad \aver{\tilde{J}_{l}^{2}}
\longrightarrow \tfrac 1 {4j^{2}} \aver{\tilde{J}_{l}^{2}} .
\end{equation}
is an entanglement condition for spin-$j$ particles.
Any quantum state that violates it is entangled.

{\it Proof.}
Let us consider a product state of $N$ spin-$j$ particles
\begin{equation}\label{prodj}
\rho_{j}= \bigotimes_{n} \rho_{j}^{(n)}
\end{equation}
and define the quantities $r^{(n)}_{l}=\frac{1}{j}\aver{j_{l}^{(n)}}.$ Then the first and modified second moments of the collective
spin can be rewritten in terms of those quantities as
\begin{equation}
\frac{\aver{J_{l}}}{2j}= \tfrac{1}{2}\sum_{n} r_{l}^{(n)}, \quad \quad \frac{\aver{\tilde{J}_{l}^{2}}}{4j^2} = \tfrac{1}{4} \sum_{n \neq m} r_{l}^{(n)}r_{l}^{(m)} .
\end{equation}
For the length of the single-particle Bloch vectors we have the constraints
\begin{equation}\label{constrj}
0\le \sum_l (r_{l}^{(n)})^{2} \leq 1.
\end{equation}
Both the lower and the upper bound are sharp, and these are the only constraints for physical states for every $j$ \cite{Constraint}. 
Thus, the set of allowed values for  $\big\{ \tfrac 1 {2j} \aver{J_{l}} \big\}_{l=x,y,z}$ 
and $\big\{\tfrac 1 {4j^{2}} \aver{\tilde{J}_{l}^{2}} \big\}_{l=x,y,z} $ for product states of the form Eq.~(\ref{prodj})
are independent from $j.$ This is also true for separable states since separable states are mixtures of product states. 
Let us now consider the range of 
\begin{align}\label{f}
f\big(\big\{ \tfrac 1 {2j} \aver{J_{l}} \big\}, \big\{\tfrac 1 {4j^{2}} \aver{\tilde{J}_{l}^{2}} \big\} \big)
\end{align}
for separable states. We have seen that the set of allowed values for the arguments of the function
in Eq.~(\ref{f}) for separable states is independent of $j.$ Thus, the range of Eq.~(\ref{f}) for separable states is also independent of $j.$ 
Hence the statement of Observation 5 follows \cite{concavity}. \qed\\

Note that the complete set of optimal \SSIs (\ref{spinjssi}) for $j>\frac{1}{2}$ can be obtained from the complete set for the spin-$\frac{1}{2}$ case presented in Ref.~\cite{TK07}
using Observation 5. 

Next, we will transform the spin-squeezing parameter $\xi_{{\rm s},j}$ to higher spins.

\observation{6}Based on Observation 5, the original spin-squeezing parameter defined in 
Eq.~(\ref{motherofallspinsqueezinginequalities}) for spin-$\frac{1}{2}$ particles
is transformed into the spin-squeezing parameter Eq.~(\ref{transosp1}) for spin-$j$ particles.

{\it Proof.} Let us first write down the entanglement condition for spin-$\frac{1}{2}$ particles based on the spin-squeezing parameter (\ref{motherofallspinsqueezinginequalities})
 in terms of the modified variance as
\begin{equation}
\xi_{{\rm s}}^2\equiv N \dfrac{( \tilde{\Delta} J_{x})^{2}+\frac{N}{4}}{\aver{J_{y}}^{2}+\aver{J_{z}}^{2}}\ge 1.
\end{equation}
Then, we use Observation 5 to obtain 
\begin{equation}\label{xisjineq}
\xi_{{\rm s},j}^2\equiv N \dfrac{( \tilde{\Delta} J_{x})^{2}+Nj^2}{\aver{J_{y}}^{2}+\aver{J_{z}}^{2}}\ge 1.
\end{equation}
\qed

It is instructive to rewrite 
Eq.~(\ref{xisjineq}) as
\begin{equation}\label{transosp}
\xi_{{\rm s},j}^2 \equiv N  \dfrac{(\Delta J_{x})^{2}}{\aver{J_{y}}^{2}+\aver{J_{z}}^{2}} 
+ N \dfrac{\sum_{n}[j^2-\aver{(j_{x}^{(n)})^{2}}]}{\aver{J_{y}}^{2}+\aver{J_{z}}^{2}} 
\geq  1 .
\end{equation}
Equation~(\ref{transosp}) can be further reformulated such that the second term depends only on the average single-particle density matrix, $\rho_{{\rm av1}},$ as
\begin{equation}\label{transosp111}
\xi_{{\rm s},j}^2=N \dfrac{(\Delta J_{x})^{2}}{\aver{J_{y}}^{2}+\aver{J_{z}}^{2}} 
+ \dfrac{j^{2}-\aver{j_{x}^{2}}_{\rm av1}}{\aver{j_{y}}_{\rm av1}^{2}+\aver{j_{z}}_{\rm av1}^{2}},
\end{equation}
where 
\begin{equation}
\label{rho1}
\rho_{{\rm av1}}:= \tfrac 1 {N} \sum_{n} \rho_{n}\equiv {\rm Tr}_2(\rho_{{\rm av2}}),
\end{equation}
and $\rho_{n}$ is the single-particle reduced density matrix for the $n^{\rm th}$ particle. Thus, in Eq.~(\ref{transosp111}) we wrote down the new
spin-squeezing parameter $\xi_{{\rm s},j}^2$ as the sum of the original parameter  $\xi_{{\rm s}}^2$ given in
Eq.~(\ref{motherofallspinsqueezinginequalities})
and 
a second term that depends only
on single particle observables and is related to single particle spin squeezing.
For $j=\frac{1}{2}$, this second term in Eq.~(\ref{transosp111}) is zero. 
For $j>\frac{1}{2},$ it is nonnegative. Hence, for $j>\frac{1}{2}$ there are states that violate Eq.~(\ref{motherofallspinsqueezinginequalities}), but
do not violate 
$\xi_{{\rm s},j}^2\ge 1.$
This is shown in a simple example with qutrits. 

{\bf Example 1.}---Let us consider a multi-particle state of the form
\begin{equation}
\vert\Psi(\alpha)\rangle=\left(\sqrt{\alpha}\vert1\rangle+
\sqrt{1-\alpha}\vert0\rangle\right)^{\otimes N} \label{eq:phialpha}
\end{equation}
for $j=1.$ For $\alpha>0.5$ and for any $N\ge1,$ the original spin-squeezing inequality (\ref{motherofallspinsqueezinginequalities})
is violated by the state \eqref{eq:phialpha}. On the other hand, no separable state can violate 
$\xi_{{\rm s},j}^2\ge 1,$
thus, it is the correct formulation of the original spin-squeezing inequality for $j>\frac{1}{2}.$ 

There is another interpretation on how to use the original spin-squeezing inequality~(\ref{motherofallspinsqueezinginequalities})
for the $j>\frac{1}{2}$ case. Equation~(\ref{motherofallspinsqueezinginequalities}) is inherently for ensembles of spin-$\frac{1}{2}$ particles. When used for higher spins, $N$ should be the number of spin-$\frac{1}{2}$ constituents rather than the number of spin-$j$ particles. Then, Eq.~(\ref{motherofallspinsqueezinginequalities}) detects entanglement between the spin-$\frac{1}{2}$ constituents of the particles, and cannot distinguish between entanglement among the spin-$j$ particles and entanglement within the spin-$j$ particles  \cite{F08}.

\observation{7}The optimal spin-squeezing inequality with three variances, Eq.~(\ref{isoin}), and the one with one variance, Eq.~(\ref{betosp}), for $(k,l,m)=(x,y,z)$ are strictly stronger than the spin-squeezing inequality  $\xi_{{\rm s},j}^2\ge 1$ [$\xi_{{\rm s},j}$ is defined in Eq.~(\ref{transosp1})],  since they
detect strictly more states.

{\it Proof.}
To see this, let us rewrite Eq.~(\ref{betosp}) for the particular choice of coordinate axes as
\begin{equation}
(N-1) \left[ (\tilde{\Delta} J_{x})^{2}+Nj^2\right] \geq \aver{\tilde{J}_{y}^{2}}+\aver{\tilde{J}_{z}^{2}}. \label{DJ}
\end{equation}
Then, from Eqs.~(\ref{isoin})  and (\ref{J2xyzJtilde2xyz}) follows
\begin{equation}
 \aver{\tilde{J}_{y}^{2}}+\aver{\tilde{J}_{z}^{2}}\ge-Nj^2+\ex{J_y}^2+\ex{J_z}^2-(\tilde{\Delta} J_{x})^{2}. 
 \label{DJ2}
\end{equation}
Clearly, the left-hand-side of Eq.~(\ref{DJ}) is not smaller than the right-hand side of Eq.~(\ref{DJ2}).
Hence, the condition $\xi_{{\rm s},j}^2\ge 1$ can be obtained.

So far we have shown that all quantum states detected by the criterion $\xi_{{\rm s},j}^2\ge 1$ are also detected by Eq.~(\ref{isoin})  or by Eq.~(\ref{betosp}) for $(k,l,m)=(x,y,z).$ 
We have now to present a quantum state that is detected by Eq.~(\ref{isoin})  or by Eq.~(\ref{betosp}) but not detected by the condition
$\xi_{{\rm s},j}^2\ge 1.$ Such states are the many-body singlet states or the symmetric Dicke states (\ref{DNj}). \qed
 
 Finally, note that the original spin-squeezing parameter $\xi_{\rm s}^2$ can also 
 be generalized to higher spins without introducing the modified
 quantities, however, in this case the bounds must be obtained
 numerically \cite{SM01}.
 
 \subsection{Other spin-$\frac{1}{2}$ criteria transformed to higher spins}
 
 In this section, we transform two generalized spin-squeezing criteria for spin$-\frac{1}{2}$ particles 
 found in the literature to criteria for higher spins.
 
 First, let us consider the criterion of Refs.~\cite{KC05,KC06}, which is valid for
multiqubit systems.
It can be rewritten in terms of the expectation values and the modified second moments as
\begin{equation}\label{kc05e}
\sqrt{\left(\aver{\tilde{J}_{l}^{2}}+\aver{\tilde{J}_{k}^{2}}\right)^{2}+(N-1)^{2}\aver{J_{n}}^{2}} -
\aver{\tilde{J}_{n}^{2}} \leq \tfrac{N(N-1)} 4.
\end{equation}
Equation~(\ref{kc05e}) is violated for some choice of the coordinate axes if the average reduced 
two-particle state is entangled \cite{Vidal}.

\observation{8}Using Observation 5, Eq.~(\ref{kc05e}) can be transformed to a system of spin-$j$ particles as
\begin{eqnarray}\label{qudkcl}
&&\sqrt{\left(\aver{\tilde{J}_{l}^{2}}+\aver{\tilde{J}_{k}^{2}}\right)^{2}+4(N-1)^{2}j^{2}\aver{J_{n}}^{2}} -
\aver{\tilde{J}_{n}^{2}}\;\;\;\;\;\;\;\;\;\;\;\;\;\;\nonumber\\
&&\;\;\;\;\;\;\;\;\;\;\;\;\;\;\;\;\;\;\;\;\;\;\;\;\;\;\;\;\;\;\;\;\;\;\;\;\;\;\;\;\;\;\;\;\;\;\;\;\;\;\;\;\;\;\leq N(N-1)j^{2}.
\end{eqnarray}

As a second example, let us consider now the entanglement condition based on the planar squeezing inequality \cite{HP11} 
for $j=\frac{1}{2}$ [i.e., Eq.~(\ref{planar}) with $C_j=\frac{1}{4}$].

\observation{9}Using Observation 5, the planar squeezing criterion can be transformed to
particles with $j>\frac{1}{2}$ as
\begin{equation}
(\tilde{\Delta} J_x)^2+(\tilde{\Delta} J_y)^2 \geq - Nj^2. \label{planarjj}
\end{equation}

It is instructive to compare Eq.~(\ref{planarjj}) to the planar spin-squeezing inequality Eq.~(\ref{planar}). 
Note again that Eq.~(\ref{planarjj}) is analytical for any $j,$ 
while Eq.~(\ref{planar}) is based on numerics.

\section{A stronger alternative of the original spin-squeezing parameter}
 \label{sec_a_stronger}
 
In this section, we show that the spin-squeezing parameter $\xi_{\spsubs}^2$ given in
 Eq.~(\ref{transosp11}), based on the optimal spin-squeezing inequality (\ref{betosp}),
is stronger than 
$\xi_{{\rm s},j}^2$  [Eq.~(\ref{transosp1})].
 In particular, it not only detects almost completely polarized spin-squeezed quantum states, 
 but also quantum states for which $\langle J_l\rangle=0$ for $l=x,y,z,$ e.g., Dicke states. 
 
How can one obtain a spin-squeezing parameter based on an entanglement condition given as an inequality?  
We will use the most straightforward way and 
divide the right-hand side of the inequality by the left-hand side, after some rearrangement of the terms.
After completing our calculations, we became aware that 
the parameter (\ref{transosp11bbb}) has appeared in Ref.~\cite{MGD12}.
It was obtained in the way described above from one of the optimal spin-squeezing inequalities for the spin-$\frac{1}{2}$ case 
[i.e., Eq.~(\ref{betosp}) with $j=\frac{1}{2}$] given in Ref.~\cite{TK07}.
 It was used to study the entanglement dynamics in the modified Lipkin-Meshkov-Glick model and its time evolution was found to be similar to the time evolution of $\xi_{\rm s}^2.$ Reference~\cite{MGD12} also describes a phase space method for the efficient 
calculation of the spin-squeezing parameters for large systems \cite{Related}. 
 
Next, we show explicitly the relation between the spin-squeezing parameter  Eq.~(\ref{transosp11})  and the corresponding optimal spin-squeezing inequality (\ref{betosp}).
Then, we prove important properties of the parameter.

\observation{10}A spin-squeezing parameter $\xi_{\spsubs}^2$ based on the optimal spin-squeezing inequality with one variance, Eq.~(\ref{betosp}), can be defined as given in Eq.~(\ref{transosp11}).
Equation~(\ref{betosp}) for $(k,l,m)=(x,y,z)$ is violated if and only if $\xi_{\spsubs}^2<1.$ 

{\it Proof. } Equation~(\ref{betosp}) can be rewritten as
\begin{equation}\label{param}
\aver{\tilde{J}_{l}^{2}} + \aver{\tilde{J}_{m}^{2}} \leq (N-1) [ (\tilde{\Delta} J_{k})^{2}+Nj^2].
\end{equation}
The spin-squeezing parameter Eq.~(\ref{transosp11})  can be obtained after dividing the right-hand side of Eq.~(\ref{param}) 
by its left-hand side. Such a derivation is valid only  if the left-hand side of Eq.~(\ref{param}) is positive.
Straightforward calculations show that if the left-hand side of Eq.~(\ref{param}) is nonpositive then Eq.~(\ref{betosp})  cannot be violated for $(k,l,m)=(x,y,z).$ \qed
 
We will now show that $\xi_{\spsubs}^2$ 
 is comparable to the original spin-squeezing parameter $\xi_{\rm s}^2.$
 
\observation{11}For large $N$  and $\xi_{{\rm s},j}^{2}<1,$  the spin-squeezing parameter $\xi_{\spsubs}^2$ is smaller than $\xi_{{\rm s},j}^{2},$ i.e., Eq.~(\ref{ineqso1}) holds.
Thus all states detected by $\xi_{{\rm s},j}^{2}$ are also detected by $\xi_{\spsubs}^{2}$ and the squeezing parameter $\xi_{\spsubs}^{2}$ is even more sensitive. 

{\it Proof.} The basic idea of the proof is that  for large $N$ the parameter $\xi_{\spsubs}^{2}$ defined in Eq.~(\ref{transosp11}) can be obtained from $\xi_{{\rm s},j}^{2}$ given in Eq.~(\ref{transosp1}) by replacing $\langle J_{l}\rangle^{2}$ with $\langle\tilde{J}_{l}^{2}\rangle$ for $l=y,z.$ Knowing that 
\begin{eqnarray}
\langle\tilde{J}_{l}^{2}\rangle\approx \langle {J}_{l}^{2}\rangle \ge \langle {J}_{l}\rangle^{2}
\end{eqnarray}
proves the claim.

We will now present a formal derivation. Straightforward algebra leads from Eq.~(\ref{transosp1}) to
\begin{eqnarray}\label{xisss}
\xi_{{\rm s},j}^{2}
&=&N\frac{[(\tilde{\Delta}J_{x})^{2}+Nj^{2}]-j(j+1)\xi_{{\rm s},j}^{2}}{\langle J_{y}\rangle^{2}+\langle J_{z}\rangle^{2}-Nj(j+1)}.
\end{eqnarray}
Let us consider first  the case when the denominator of  Eq.~(\ref{xisss}) is positive.
Then, we need the relation between the expectation values and the second moments
\begin{equation} \label{J22}
\langle{J}_{l}^{2}\rangle \ge\langle J_{l}\rangle^{2},
\end{equation}
and the relation between the modified second moments and the true second moments 
\begin{eqnarray}\label{Jyz}
\langle\tilde{J}_{y}^{2}\rangle+\langle\tilde{J}_{z}^{2}\rangle\ge\langle J_{y}^2\rangle+\langle J_{z}^2\rangle-Nj(j+1).
\end{eqnarray}
Equation~(\ref{Jyz}) can be easily derived from Eq.~(\ref{J2xyzJtilde2xyz}).
Based on Eqs.~(\ref{J22}) and (\ref{Jyz}), we obtain an inequality for the usual spin-squeezing parameter
\begin{eqnarray}
\xi_{{\rm s},j}^{2} &\ge&
(N-1)\frac{[(\tilde{\Delta}J_{x})^{2}+Nj^{2}]-j(j+1)\xi_{{\rm s},j}^{2}}
{\langle\tilde{J}_{y}^{2}\rangle+\langle\tilde{J}_{z}^{2}\rangle}.\nonumber\\\label{qqqqqq}
\end{eqnarray}
Let us compare Eq.~(\ref{qqqqqq})  with the fraction in Eq.~(\ref{transosp11}). 
One can see that the only difference is 
the  $j(j+1)\xi_{{\rm s},j}^{2}$ term in the numerator of Eq.~(\ref{qqqqqq}).
If $\xi_{{\rm s},j}^{2}<1$  then for large $N$  the first term in the numerator in Eq.~(\ref{qqqqqq}) is much larger than the second one 
\begin{eqnarray}\label{qqq}
[(\tilde{\Delta}J_{x})^{2}+Nj^{2}]\gg j(j+1)\xi_{{\rm s},j}^{2}.
\end{eqnarray}
This can be seen noting that $(\tilde{\Delta}J_{x})^{2}+Nj^{2}\ge \va{J_x}$ holds and for large particle numbers the variance of an angular momentum component is, in practice, much larger than $\sim1.$
Thus, for large particle numbers the right-hand side of Eq.~(\ref{qqqqqq}) equals $\xi_{\spsubs}^2.$

Finally, note that if the denominator of  Eq.~(\ref{xisss}) is nonpositive then 
the condition $\xi_{{\rm s},j}^2<1$ can be satisfied only
if $(\tilde{\Delta}J_{x})^{2}+Nj^{2} \le j(j+1)$ which would be possible if $(\tilde{\Delta}J_{x})^{2}\sim1$ and hence is not realistic for large particle numbers.
$\qed$

Observation 11 is valid only for large particle numbers. For small particles numbers,
there are quantum states that are detected by the original spin-squeezing parameter generalized for arbitrary spin, Eq.~(\ref{transosp1}), but not detected by the spin-squeezing parameter $\xi_{\spsubs}^2$ defined in Eq.~(\ref{transosp11}).
For instance, such a state is a ground state of the five-qubit Hamiltonian 
\begin{eqnarray} \label{H5}
H_5=J_x^2+\tfrac{1}{4} J_z^2+\tfrac{3}{4}J_z.
\end{eqnarray}
The Hamiltonian (\ref{H5}) has a four dimensional subspace of ground states.
Any state in this subspace has $\xi_{\rm s}^2=0.97$ while $\xi_{\spsubs}^2=1.29.$
Due to Observation 7, these states must violate the
optimal spin-squeezing inequality with three variances, Eq.~(\ref{isoin}),
which can be verified by direct calculation. 

It is instructive to see, how the spin-squeezing parameter $\xi_{\spsubs}^2$
behaves for an ensemble of particles almost fully polarized in the $z$ direction. 
For a fully polarized ensemble, the first and second moments of the angular momentum components are
\begin{equation} \label{fully}
\begin{array}{ll}
\exs{J_x^2}=\exs{J_y^2}=\tfrac{1}{2}Nj,&\;\;\;\;\;\;\exs{J_z^2}=N^2j^2,\\
\exs{J_x}=\exs{J_y}=0,&\;\;\;\;\;\;\exs{J_z}=Nj.
\end{array}
\end{equation} 
Based on these, we obtain the following formulas, which are approximately valid
for almost fully polarized ensembles
\begin{subequations} \label{compare}
\begin{eqnarray} \label{approx1}
\xi_{\spsubs}^2 &\approx& \frac{(\tilde{\Delta} J_{x})^{2}+Nj^2}{Nj^2+
\frac{1}{(N-1)}[\exs{J_y^2}+\sum_n \exs{(j_x^{(n)})^2}-Nj]},\nonumber\\
\\
\xi_{{\rm s},j}^2  &\approx& \frac{(\tilde{\Delta} J_{x})^{2}+Nj^2}{Nj^2}.
\end{eqnarray}
\end{subequations}
In Eq.~(\ref{compare}), we substituted the value for completely polarized states for $\exs{J_z}$ and $\exs{J_z^2}.$
We also used Eq.~(\ref{jxyz2}) to eliminate $j_y$ and $j_z$ from  Eq.~(\ref{approx1}).
The second term in the denominator of Eq.~(\ref{approx1})
is negligible compared to the first  term which is $\propto N.$
Hence,  the two spin-squeezing parameters are approximately equal
\begin{equation}\label{xiso}
\xi_{\spsubs}^2\approx \xi_{{\rm s},j}^2.
\end{equation}
Thus, the spin-squeezing parameter  $\xi_{\spsubs}^2$ detects the fully polarized
entangled states detected by
$\xi_{{\rm s},j}^2.$

In practical situations, the almost completely polarized state is mixed with noise.
Next, we will discuss  noisy spin-squeezed states.

\observation{12}The spin-squeezing parameter $\xi_{\spsubs}^2$  is much
more efficient than $\xi_{{\rm s},j}^2$ in detecting almost completely polarized spin-squeezed states
mixed with white noise.

{\it Proof.} Let us consider a state  $\varrho$  that is almost completely polarized in the $z$ direction  and spin-squeezed in the $x$ direction. 
After mixing $\varrho$ with white noise, we obtain
\begin{equation}
\varrho_{{\rm noisy}}(p_{{\rm n}})=(1-p_{{\rm n}})\varrho+p_{{\rm n}}\rho_{{\rm cm}} ,
\end{equation}
where $p_{{\rm n}}$ is the ratio of noise and $\rho_{{\rm cm}}$ is defined in Eq.~(\ref{cm}). 
Then, using that we have $\exs{J_x}=0,$ 
straightforward calculations show that the original spin-squeezing parameter increases more
\begin{equation}\label{noisypar}
{\xi_{{\rm s},j,{\rm noisy}}^2}
=\frac{1}{(1-p_{\rm n})}{\xi_{{\rm s},j}^2} +\frac{p_{\rm n}}{(1-p_{\rm n})^2}\frac{N^2j^2}{(\langle J_y \rangle^2+\langle J_z \rangle^2)},
\end{equation}
than our alternative spin-squeezing parameter
\begin{equation}\label{noisypar2}
\xi_{{\spsubs},{\rm noisy}}^2=
{\xi_{{\spsubs}}^2}+\frac{p_{\rm n}}{1-p_{\rm n}}\frac{N(N-1)j^2}{(\langle\tilde{J}_{y}^{2}\rangle+\langle\tilde{J}_{z}^{2}\rangle)}.
\end{equation}
Since Eq.~(\ref{xiso}) and $\langle J_y \rangle^2+\langle J_z \rangle^2\approx
\langle\tilde{J}_{y}^{2}\rangle+\langle\tilde{J}_{z}^{2}\rangle$
hold for almost fully polarized spin-squeezed states and for large particle numbers,  we obtain
\begin{equation}
\xi_{{\spsubs},{\rm noisy}}^2\approx {\xi_{{\rm s},j,{\rm noisy}}^2} (1-p_{\rm n}).  
\end{equation}
This proves our claim.
$\qed$

Besides almost completely polarized states, our spin-squeezing parameter 
$\xi_{\spsubs}^2$ can also detect the entanglement of unpolarized states.
This is due to the fact that it is defined in Eq.~(\ref{transosp11}) based on the spin-squeezing inequality~(\ref{betosp}), which can be used to detect the symmetric Dicke state $\vert D_{N,j}\rangle,$ given in Eq.~(\ref{DNj}).
Such states have $\exs{J_l}=0$ for $l=x,y,z,$ and thus they are not detected by $\xi_{{\rm s},j}^2$
\cite{xis_Dicke}.
We will now analyze how it is possible that Eq.~(\ref{transosp11}) can be used to detect both usual spin-squeezed states with a large polarization $\vert \vec J \vert$ and states 
 with $\vec J=0.$
  
 For that, let us rewrite Eq.~(\ref{transosp11}) such that
 the denominator contains both variances of the spin components and their expectation values
 \begin{equation}
\xi_{\spsubs}^2=(N-1)\dfrac{(\tilde{\Delta} J_{x})^{2}+Nj^2}{( \tilde{\Delta}{J}_{y})^2+(\tilde{\Delta} {J}_{z})^2+\exs{J_y}^2+\exs{J_z}^2}. 
\end{equation}
Thus, the states detected by $\xi_{\spsubs}^2<1$ have to have a small variance of a spin component in some direction. Then, in the orthogonal directions either they have to have a large 
spin component or a large variance of one of those spin components. 

\begin{figure}
\centerline{ \epsfxsize3in
\epsffile{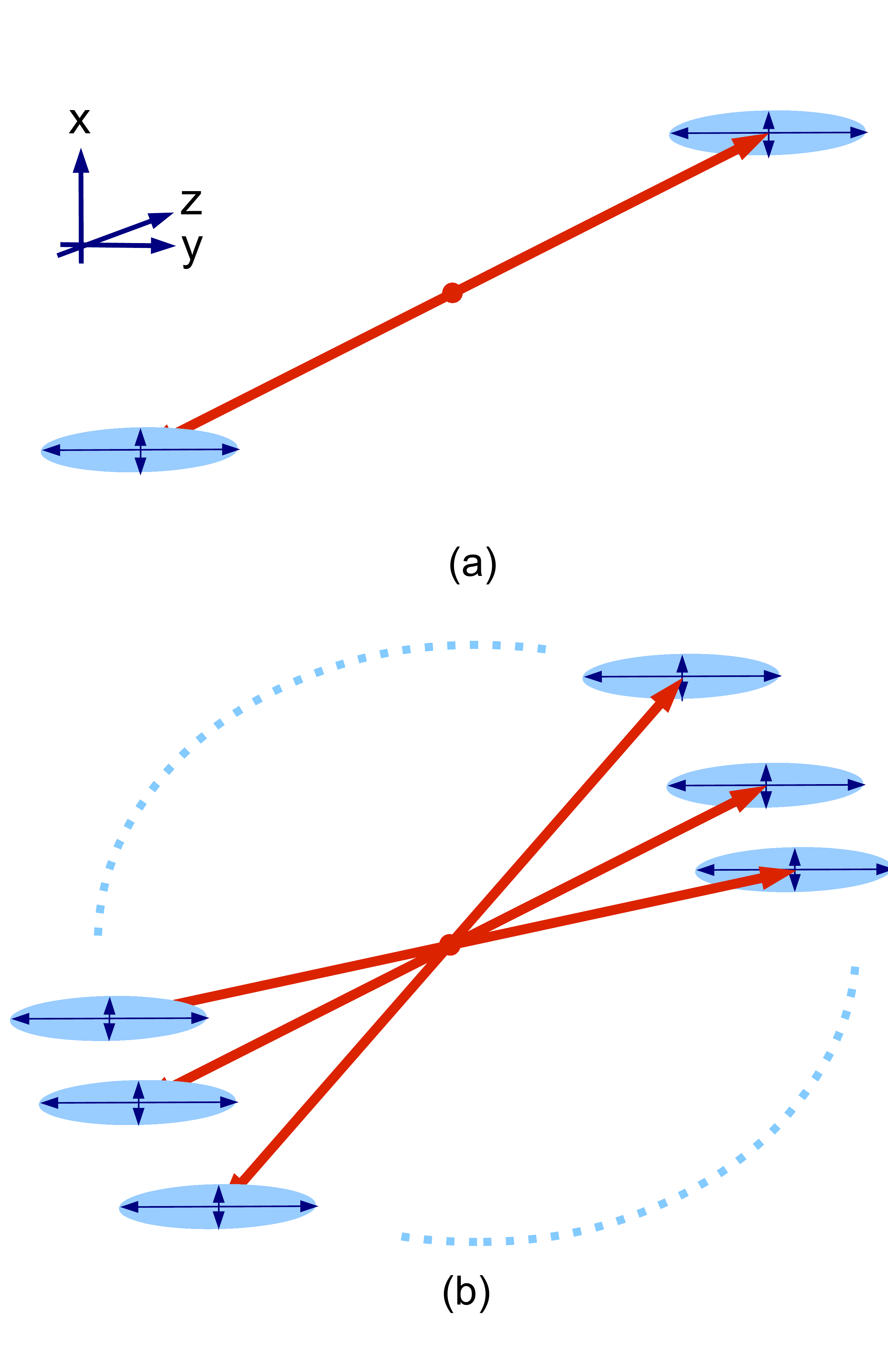}
} 
\caption{(Color online) Different types of spin-squeezed states violating the criterion based on the spin-squeezing parameter $\xi_{\spsubs}^2$ [Eq.~(\ref{transosp11})] having ${\vec J}=0.$ (a) Mixture of two almost completely polarized spin-squeezed states pointing into opposite directions. (b) Mixture of several almost completely polarized spin-squeezed states. The original spin-squeezing inequality based on the spin-squeezing parameter and its generalization for arbitrary spin $j,$ $\xi_{{\rm s},j}^2$ [Eq.~(\ref{transosp1})] cannot detect these states since for these states the mean spin is zero. } \label{spisqfig2} 
\end{figure}

Let us see now an application of the ideas above.

\observation{13}Consider a set of quantum states 
$\varrho_k$ such that (i) $\varrho_k$ are 
all detected as entangled by the spin-squeezing parameter $\xi_{\spsubs}^2$ and 
(ii)
\begin{equation}\label{sss1}
(\tilde{\Delta}J_x)^2_{\varrho_k}=(\tilde{\Delta}J_x)^2_\varrho,
\end{equation}
for all $k.$ Then, their mixture 
\begin{equation}\label{sss2}
\varrho=\sum_k p_k\varrho_k
\end{equation}
for $p_k>0$ and $\sum_k p_k=1$ is always detected as entangled by $\xi_{\spsubs}^2.$
This is not the case for 
the spin-squeezing parameter $\xi_{{\rm s},j}^2$  defined in Eq.~(\ref{transosp1}).
For an illustration, see Fig.~\ref{spisqfig2}.

{\it Proof.} The observation can be proved by straightforward substitution of 
Eqs.~(\ref{sss1}) and (\ref{sss2}) into  Eq.~(\ref{transosp11}). $\qed$

Following Observation 13, let us consider a spin-squeezed state $\varrho_{\rm ss}$ of many particles that is almost completely polarized in the $z$ direction and spin squeezed along the $x$ direction. Such a state is detected by the spin-squeezing parameter $\xi_{\spsubs}^2$ defined in Eq.~(\ref{transosp1}) and also by the parameter $\xi_{{\rm s},j}^2.$
Due to Observation 13, the following state is also detected by $\xi_{\spsubs}^2$ 
\begin{equation}
\varrho_{\rm ss,rot}=\frac{1}{2\pi}\int_{0}^{2\pi} {\rm d}\phi  e^{-iJ_x\phi} \varrho_{\rm ss} e^{+iJ_x\phi}. \label{ss2}
\end{equation}
The quantum state (\ref{ss2}) has $\ex{J_l}=0$ for $l=x,y,z,$ a large value for $( \tilde{\Delta}{J}_{y})^{2}+(\tilde{\Delta} {J}_{z})^{2}$ and a small value for $( \tilde{\Delta}{J}_{x})^{2}.$ From the point of view of collective observables, the state (\ref{ss2}) is similar to the symmetric Dicke state given in Eq.~(\ref{DNj}). Such a state is clearly not detected as entangled by the parameter $\xi_{{\rm s},j}^2.$ The state $ \varrho_{\rm ss,rot}$  is depicted in Fig.~\ref{spisqfig2}(b).

Finally, note that spin-squeezing parameters can be defined based on the optimal spin squeezing inequality with three variances  Eq.~(\ref{isoin}) as \cite{TM10}
\begin{equation}
\xi_{\rm singlet}^2=\dfrac{\sum_l ({\Delta} J_{l})^{2}}{Nj}. 
\end{equation}
For a pure state, the quantity $N\xi_{\rm singlet}^2$ gives an upper bound on the number of particles not entangled with other particles \cite{TM10,singlet}. $\xi_{\rm singlet}^2$ can also be interpreted through connections to robustness measures \cite{CP11}.

It is also possible to define a spin squeezing parameter based on the inequality with two variances Eq.~(\ref{twovar}) as
\begin{equation} \label{pl}
\xi_{\rm planar\;squeezing}^2=(N-1) \dfrac{(\tilde{\Delta} J_{x})^{2}+(\tilde{\Delta} J_{y})^{2}+Nj^2}{\aver{\tilde{J}_{z}^{2}}}.
\end{equation}
If the parameter (\ref{pl}) is smaller than 1, and the denominator is positive then the state is entangled.
Equation~(\ref{pl})  can be used to characterize planar squeezing.
  
\section{Further considerations} \label{further}

Next, we will discuss several issues connected mostly to practical 
aspects of using the spin squeezing inequalities for entanglement detection.

\subsection{The nematic tensor and single-particle spin squeezing}\label{singlepart_spinsq}

In this section we discuss that single-particle spin squeezing becomes possible for particles with $j>\frac{1}{2},$ and it is characterised by the local second moments.

As mentioned in the introduction, for spin-$\frac{1}{2}$ particles 
the local second moments
$\exs{\sum_n(j_l^{(n)})^2}$ are constants. For $j>\frac{1}{2},$ 
the local second moments
are not constants any more.  In order to characterize the collective local second moments
in any direction, we introduce the following matrix
\begin{gather}
\nmatrix_{kl}:=\tfrac 1 {N} \sum_n \bigg(\tfrac{1}{2}\aver{j_k^{(n)} j_l^{(n)} +
j_l^{(n)} j_k^{(n)}}-Q_0\delta_{kl}\bigg),\label{Dkl}
\end{gather}
where for convenience we define
\begin{equation}Q_0:=\tfrac{j(j+1)}{3}.
\end{equation} 
The  traceless $\nmatrix$  matrix is the rank-2 quadrupole or nematic tensor \cite{CM04,AG84,CC91,HG12,SL10,YU13,LC}.
It depends only on the average single particle density matrix thus 
it can be rewritten as
\begin{gather}
\nmatrix_{kl}=\bigg(\tfrac{1}{2}\aver{j_k j_l +j_l j_k}_{\rm av1}-Q_0\delta_{kl}\bigg),\label{Dkl_rho1}
\end{gather}
where the average singe-body density matrix $\varrho_{\rm av1}$ is defined in Eq.~(\ref{rho1}).
The  second moment of any angular momentum component can be obtained as
\begin{gather}
\exs{j^2_{\vec n}}_{\rm av1}={\vec n}^T (\nmatrix+Q_0\openone) \vec n,
\end{gather}
where the unit vector  $\vec n$ describes the direction of the component.

The matrix $\nmatrix,$ together with the average single particle spin
\begin{equation}\exs{\vec{j}}=\tfrac{1}{N}(\exs{J_x}, \exs{J_y}, \exs{J_z}),
\end{equation}
contains all the information to calculate the single-particle average spin squeezing parameter
\begin{equation}
\xi_{{\rm s},j,{\rm av1}}^2=2j\frac{(\Delta j_{\vec n})_{\rm av1}^2}{\exs{j_{\vec n \perp1}}_{\rm av1}^2+\exs{j_{\vec n \perp2}}_{\rm av1}^2},
\end{equation}
where $\vec n$ is some direction, and  $j_{\vec n \perp k}$ are two directions perpendicular to $\vec n$ and to each other. If $\xi_{{\rm s},j,{\rm av1}}^2<1$ then there is entanglement
between the $2j$ spin-$\frac{1}{2}$ constituents within the average single-particle state \cite{spin12}.
For $j>\frac{1}{2},$ it is possible to obtain spin squeezing within the particles, which can lead to improvement in metrological applications, but does not involve interparticle entanglement 
\cite{F08,NT12}. 
 
In Eq.~(\ref{tildequant}), we defined the modified second moments and modified variances
that do not contain the local second moments. 
Thus, our inequalities for the spin-$j$ particles can be interpreted as entanglement conditions that
separate the entanglement between the spin-$\frac{1}{2}$ constituents of the spin-$j$ particles and entanglement between the spin-$j$ particles. 
Our inequalities detect only spin squeezing due to interparticle entanglement.

\subsection{Coordinate system independent form of the spin squeezing inequalities}\label{indform}

In this section, we show how to write down the optimal spin squeezing inequalities for a general $j$
in a form that is independent from the choice of the coordinate axes. Such a form of our inequalities is very useful, as one does not have to look for the optimal choice of the coordinate axes for the spin squeezing inequalities to detect a given quantum state as entangled.

First, we define the quantities that are necessary to characterize the second moments
and covariances of collective angular momentum components \cite{sum}
\begin{gather}
C_{kl}:=\tfrac 1 2 \aver{J_k J_l+J_l J_k} , \notag \\
\gamma_{kl}: = C_{kl} -\aver{J_k}\aver{J_l} .
\end{gather}
The matrices $C$ and $\gamma$ have already been defined for the 
optimal spin squeezing inequalities for  $j=\frac{1}{2}$ \cite{TK07,CGamma}.
For the $j>\frac{1}{2}$ case,  we also need the nematic matrix $\nmatrix$ given in Eq.~(\ref{Dkl}) to characterize the local second moments of the angular momentum coordinates.

Based on these, we define the matrix that will play a central role in our entanglement conditions
\begin{equation}\label{XX}
\mathfrak{X} := (N-1)\gamma +C -N^2\nmatrix .
\end{equation}
The matrix $\mathfrak{X}$ has also been introduced for spin-$\frac{1}{2}$ particles in Ref.~\cite{TK07}.
 For such systems $\nmatrix=0\cdot\openone$ and $\mathfrak{X} = (N-1)\gamma +C,$
 which agrees with the definition in Ref.~\cite{TK07}.
 
 We can now present our coordinate 
 system independent entanglement criteria. 

\observation{14}The coordinate system independent form of the
optimal spin squeezing inequalities for spin-$j$ particles is
\begin{subequations}\label{spinjssi_ci}
\begin{eqnarray}
{\rm Tr}(C) &\leq& Nj(Nj+1) ,  \label{symmsatin_ci} \\
{\rm Tr}(\gamma)  &\geq& Nj ,  \label{isoin_ci}  \\
  \lambda_{\min}(\mathfrak{X}) &\geq& \trace(C) -Nj(Nj+1)+N^2Q_0,  \label{betosp_ci}  \\
\lambda_{\max}(\mathfrak{X}) &\leq&  (N-1)\trace(\gamma)  -N(N-1)j+N^2Q_0,  \label{twovar_ci} \nonumber\\
\end{eqnarray}
\end{subequations}
where $\lambda_{\min}(A)$ and $\lambda_{\max}(A)$ are the smallest and largest eigenvalues of the matrix $A,$ respectively.

{\it Proof.} Equation~(\ref{symmsatin_ci})  can be obtained straightforwardly by 
replacing the sum of the three second moments by $\trace(C)$ on the left-hand sides of Eq.~(\ref{spinjssi}a). 
Similarly, Eq.~(\ref{isoin_ci}) can be obtained by 
replacing the sum of the three variances by 
$\trace(\gamma)$ on the left-hand side of Eq.~(\ref{spinjssi}b). 

In order to obtain Eq.~(\ref{betosp_ci}) from Eq.~(\ref{betosp}), we need to add $\exs{\tilde{J}_k^2}$ to both sides of 
Eq.~(\ref{betosp})
\begin{equation}
\aver{\tilde{J}_{k}^{2}}+\aver{\tilde{J}_{l}^{2}} + \aver{\tilde{J}_{m}^{2}}  -N(N-1)j^{2} \leq (N-1)  (\tilde{\Delta} J_{k})^{2}+\aver{\tilde{J}_{k}^{2}}.\label{betosp_plustildeJz2} 
\end{equation}
Then, we need to
write down explicitly a diagonal element of the matrix defined in 
Eq.~(\ref{XX}) with the modified second moments and variances as
\begin{equation}\label{Xkk_diagonal} 
\mathfrak{X}_{kk}=(N-1) (\tilde{\Delta }J_k)^2+\exs{\tilde{J}_k^2} + N^2Q_0,
\end{equation}
where $k\in\{x,y,z\}.$
Using Eqs.~(\ref{J2xyzJtilde2xyz}) and (\ref{Xkk_diagonal}), 
the optimal spin squeezing inequality with a single variance, Eq.~(\ref{betosp_plustildeJz2}), can be rewritten as 
\begin{equation}\label{betosp_ci2} 
\mathfrak{X}_{kk}-N^2Q_0
\ge {\rm Tr}(C) -Nj(Nj+1).
\end{equation}
$\mathfrak{X}_{kk}$ is the only quantity in Eq.~(\ref{betosp_ci2}) that depends on the choice of coordinate axes.
Equation~(\ref{betosp_ci2}) is violated for some choice of the coordinate axes, if and only if Eq.~(\ref{betosp_ci}) is violated. A similar derivation leads from Eq.~(\ref{twovar})  to Eq.~(\ref{twovar_ci}).  \qed

\subsection{Measuring the second moments of local operators}\label{addcom}

In this section, we will discuss the additional complexity arising from the need to measure 
the modified second moments of the collective angular momentum components, given in Eq.~(\ref{tildequant}), rather than the true second moments, for $j>\frac{1}{2}.$  We will show that for each inequality it is sufficient to measure at most only one of the quantities $M_l$ defined in 
Eq.~(\ref{Mvec}).

Let us now take the four inequalities in  Eq.~(\ref{spinjssi}) and examine whether they need the measurement of the modified second moments.
Two of the inequalities, namely Eqs.~(\ref{symmsatin}) and (\ref{isoin}), are 
already written in terms of the true variances and second moments. In Eq.~(\ref{betosp}), 
all the three expectation values, $M_k,$ $M_l,$ and $M_m,$ appear.
Based on Eq.~(\ref{jxyz2}), from
Eq.~(\ref{betosp}) we obtain
\begin{align}
&(N-1)  (\Delta J_{k})^{2} - NM_k
   \notag \\
&\;\;\;\;\;\;\;\;\;\;\;\;\;\;\;\;\;\;\;\;\;\;\;\;\geq -Nj(Nj+1) + \left[\aver{J_{l}^{2}} + \aver{J_{m}^{2}} \right] .  \label{othtwo2}
\end{align}
In an analogous way, we can transform Eq.~(\ref{twovar}) to
\begin{align}
&(N-1) \left[ (\Delta J_{k})^{2} + (\Delta J_{l})^{2} \right] 
\notag \\ 
  &\;\;\;\;\;\;\;\;\;\;\;\;\;\;\;\;\;\;\;\;\;\;\;\;\geq 
  N(N-1)j  + \aver{ J_{m}^{2}}-NM_m. \label{othtwo1} 
\end{align}
Note that a similar equation has already been used in Eq.~(\ref{othtwo1bbb}) to describe planar squeezing.
It can be seen explicitly that  both for Eqs.~(\ref{othtwo2}) and (\ref{othtwo1}) only
the measurement of one of the second moments of the local operators is needed.

Measuring the expectation value of the operator $\sum_n (j_l^{(n)})^2$ can be realized in two different ways:
(i) by rotating the spin by a magnetic field, and then measuring the populations of the $j_z$ eigenstates. 
Let us denote the eigenvalues of $j_z$ by $\chi_z.$
The sum of the local second moments can be obtained with the populations of the $j_z$ eigenstates, $N_{\chi_z},$ as
\begin{equation}
M_z=\sum_{\chi_z=-j,-j+1,...,j} N_{\chi_z}\chi_z^2.
\end{equation}
For spin-$1$ systems, $M_z=N_{-1}+N_{+1}=N-N_0.$

(ii) In some cold atomic systems, such operators might also be measured directly, as in such systems in the Hamiltonian a  $(j_l^{(n)})^2$ 
term coupled to the pseudo spin of the light appears \cite{EM05,NM09,KM10}. 

One might try to eliminate the need for measuring quantities of the
type $\aver{M_m}$ in Eq.~(\ref{othtwo2}) by looking for the 
minimum of  $(\Delta J_{k})^{2}$ for a given  $\left[\aver{J_{l}^{2}} + \aver{J_{m}^{2}} \right].$
This problem is very complex and possibly can only be solved numerically for large spins.
Analogously, a condition similar to Eq.~(\ref{othtwo1}), 
but without the need for measuring $\aver{M_m}$ can be obtained by looking for the 
minimum of $\left[ (\Delta J_{k})^{2} + (\Delta J_{l})^{2} \right]$  for a given 
$\aver{ J_{m}^{2}}.$

\section{Summary}

In summary, we have presented a complete set of generalized spin squeezing inequalities 
for detecting entanglement in an ensemble of spin-$j$ particles with $j>\frac{1}{2}$ based on knowing only
$\exs{J_l}$ and the modified second moments $\exs{\tilde{J}^2_l}$ for $l=x,y,z.$
We have called the inequalities optimal spin squeezing inequalities for spin-$j$ particles.
We have also presented a mapping from \SSIs valid in qubit systems to \SSIs valid in qudit systems. 
We have shown how to transform the original spin squeezing parameter to an ensemble of particles
with a spin larger than $\frac{1}{2}.$ 
We have shown that a new spin squeezing parameter based on the optimal spin squeezing inequality with a single variance 
is, for large particle numbers,  strictly stronger than the original
spin squeezing parameter and its version mapped to higher spins.
We have also examined the entanglement properties of the states detected by our
inequalities and computed the noise tolerances of our inequalities for these states.
We have also discussed how to measure the modified second moments in experiments.

In the future, it would be interesting to extend our research to entanglement conditions based on
collective observables different from angular momentum operators, with collective operators based on the $SU(d)$ generators \cite{PRL,OTHER}. Moreover, it would also be interesting 
to find entanglement conditions with the true second moments, without the need for measuring 
the modified second moments even if this involves numerical calculations rather than analytical ones.

\begin{acknowledgments}
We thank P. Hyllus for many useful discussions on spin squeezing and entanglement.
We acknowledge Z. Zimbor\'as for many interesting discussions on group theory, especially
for pointing out the importance of symmetric singlet states for particles with a higher spin.
We also thank G. Colangelo, F. Deuretzbacher, O. G\"uhne, Z. Kurucz, 
C. Klempt, M.W. Mitchell, M. Modugno, L. Santos, 
R. J. Sewell, and I. Urizar-Lanz 
for discussions. We thank the EU (ERC Starting Grant GEDENTQOPT, CHIST-ERA QUASAR)
the Spanish MINECO
(Project No. FIS2009-12773-C02-02 and No. FIS2012-36673-C03-03),
the Basque Government (Project No. IT4720-10), and
the National Research Fund of Hungary OTKA (Contract No. K83858).
\end{acknowledgments}

\appendix

\section{Proof of Eq.~(\ref{jx2Dicke})}

In this appendix, we present a proof of the formula Eq.~(\ref{jx2Dicke}) for symmetric
Dicke states with a maximal $\langle J_{x}^{2}+J_{y}^{2}+J_{z}^{2} \rangle$ and 
$\langle J_{z}\rangle=0$. For completeness, we will
consider a more general case, namely, states for which 
 $\langle J_{z}\rangle\ne 0$ is also allowed. For carrying out our calculations, 
we need to map states of $N$ spin-$j$ particles to states of $2Nj$
spin-$\frac{1}{2}$ particles. 

\observation{15}Let us consider symmetric Dicke states  of $N$ spin-$j$ particles
\begin{equation} \label{b1}
\vert Nj,\lambda_z\rangle_{j},
\end{equation}
which fulfill the eigenequations
\begin{eqnarray}\label{eigen}
(J_{x}^{2}+J_{y}^{2}+J_{z}^{2})\vert Nj,\lambda_z\rangle_{j} &=& Nj(Nj+1) \vert Nj,\lambda_z\rangle_{j},\nonumber\\
J_z \vert Nj,\lambda_z\rangle_{j} &=&\lambda_z \vert Nj,\lambda_z\rangle_{j},
\end{eqnarray}
and the subscript $j$ indicates that they are states of spin-$j$ particles. For such states,
\begin{equation}\label{jx2Dickeb}
\sum_n \aver{(j_z^{(n)})^2}= \tfrac{N(N-1)j^2}{2jN-1}+\tfrac{2j-1}{(2Nj-1)}\lambda_z^2.
\end{equation}

{\it Proof.} First note for the quantum state (\ref{b1}) $\exs{J_{x}^{2}+J_{y}^{2}+J_{z}^{2}}$
is maximal. All such states are uniquely characterised by the two eigenvalues in the
eigenequations Eq.~(\ref{eigen}). Thus, a third parameter to distinguish states
with degenerate eigenvalues is not needed.

Analogously, a symmetric Dicke state of $2Nj$
spin-$\frac{1}{2}$ particles satisfying also the property that $\exs{J_{x}^{2}+J_{y}^{2}+J_{z}^{2}}$
is maximal can be denoted as
\begin{equation} \label{b2}
\vert Nj,\lambda_z\rangle_{\frac{1}{2}},
\end{equation}
where quantum state (\ref{b2})  also fulfils the eigenequations 
Eq.~(\ref{eigen}).

The symmetric Dicke state of spin-$j$ particles, Eq.~(\ref{b1}),
can be mapped to the Dicke state of spin-$\frac{1}{2}$ particles, Eq.~(\ref{b2})
\begin{equation}
\vert Nj,\lambda_z\rangle_{j}\rightarrow\vert Nj,\lambda_z\rangle_{\frac{1}{2}}.
\end{equation}
The moments of the collective angular momentum
components, $\exs{J_l^{m}},$  are the same for the states
$\vert Nj,\lambda_z\rangle_{j}$ and $\vert Nj,\lambda_z\rangle_{\frac{1}{2}}.$
We can imagine that we represent the spin-$j$ particle as $2j$
spin-$\frac{1}{2}$ particles in a symmetric state. 
For example, the spin-$1$ state $\ket{0}$ 
is mapped to a symmetric two-qubit state
\begin{equation}
\vert 1,0\rangle_{1}\equiv \ket{0}\rightarrow\vert 1,0\rangle_{\frac{1}{2}}
\equiv \tfrac{1}{\sqrt{2}}(\ket{+\tfrac{1}{2},-\tfrac{1}{2}}+\ket{-\tfrac{1}{2},+\tfrac{1}{2}}).
\end{equation}

Operators can be mapped in an analogous way. 
The expectation value of the square of the single-particle operator for spin-$j$ particles can be expressed
with operators acting on the spin-$\frac{1}{2}$ state as 
\begin{equation}
\langle (j_{z}^{(n)})^{2}\rangle_j=\bigg\langle\bigg(\sum_{k=1}^{2j} j_z^{(n,k)}\bigg)^{2}\bigg\rangle_{\frac{1}{2}}.\label{eq:j2_2Nj}
\end{equation}
The left-hand side is an expectation value evaluated on $\vert Nj,\lambda_z\rangle_{j},$
while the right-hand side  is an expectation value evaluated on $\vert Nj,\lambda_z\rangle_{\frac{1}{2}}.$
The superscript $(n,k)$ denotes the  $k^{\rm th}$ spin-$\frac{1}{2}$ constituent of the to the $n^{\rm th}$ qudit.

In the following, we will refer only to the state with spin-$\frac{1}{2}$ particles and hence will omit the 
${\frac{1}{2}}$ subscript. Due to symmetry we can express 
the right-hand side of Eq.~(\ref{eq:j2_2Nj})
with single-particle and two-particle expectation values as
\begin{equation}\label{jz2j}
\bigg\langle\bigg(\sum_{k=1}^{2j} j_z^{(n,k)}\bigg)^{2}\bigg\rangle=
2j\langle (j_{z}^{(n,1)})^{2}\rangle+2j(2j-1)\langle j_{z}^{(n,1)}j_{z}^{(n,2)}\rangle.
\end{equation}
The single-particle second moment is
\begin{equation}\label{111}
\langle(j_{z}^{(n,1)})^{2}\rangle =\tfrac{1}{4}.
\end{equation}
Moreover, the two-body correlations can be calculated from $\langle J_{z}^{2}\rangle=\lambda_z^2$
as
\begin{equation}\label{222}
\langle j_{z}^{(n,1)}j_{z}^{(n,2)}\rangle =-\tfrac{1}{4(2Nj-1)}+\tfrac{1}{2Nj(2Nj-1)}\lambda_z^2.
\end{equation}
Substituting Eqs.~(\ref{111}) and  (\ref{222}) into Eq.~(\ref{jz2j}), Eq.~(\ref{jx2Dickeb}) follows. $\qed$


\begin{thebibliography}{99}
\bibitem{Note_tilde} Note that the quantity $\tilde{J}_{l}$ in itself is not defined, only the modified second moment and variance,
$\aver{\tilde{J}_{l}^{2}}$ and $(\tilde{\Delta}J_{l})^{2}$. These latter quantities are not the second
moment and the variance of any observable and they can also attain negative values.
\bibitem{Constraint} Note that we considered products of mixed states of the form (\ref{prodj}).
If we considered products of pure states then for $j=\frac{1}{2}$ we had $\sum_l (r_{l}^{(n)})^{2}=1$
instead of Eq.~(\ref{constrj}).
\bibitem{concavity} In Ref.~\cite{PRL}, the concavity of the function $f$ was also required. Here, we presented a proof for which concavity is not necessary.
\bibitem{Related} In Ref.~\cite{RevNori}, another parameter has been defined based on the same optimal spin squeezing inequality as $\xi_{\rm E}^2=N\va{J_x}/(\exs{\vec{J^2}}-\frac{N}{2}-\exs{J_x}^2).$ The difference between $\xi_{\rm E}^2$ and $\xi_{\spsubs}^2$ is that $\va{J_x}$ is added to both the numerator and the denominator. Due to that, if  $\xi_{\spsubs}^2<1$ then $\xi_{\rm E}^2\ge  \xi_{\spsubs}^2,$ 
 thus $\xi_{\spsubs}^2$ is more sensitive to entanglement.
 For symmetric states, $\xi_{\rm E}^2$ equals the parameter proposed  in Ref.~\cite{RP01} for detecting spin squeezing close to Dicke states. Finally, note that in Ref.~\cite{DuanPRL}, a parameter for multipartite entanglement for spin-$\frac{1}{2}$ particles has been defined, which is especially useful for Dicke states. It needs the same operators to be measured as the inequality (\ref{betosp}) for the spin-$\frac{1}{2}$ case, however, is not based on that inequality.
\bibitem{xis_Dicke} For the symmetric Dicke state~(\ref{DNj}) for $j=\frac{1}{2},$ both the numerator and the denominator of the fraction in the definition of  $\xi_{{\rm s},j},$ given in Eq.~(\ref{transosp1}), are zero. However, it is easy to see that if the state is mixed with any state that is not detected as entangled by the spin squeezing parameter $\xi_{{\rm s},j}$ then this mixture is not detected either. For $j>\frac{1}{2},$ the numerator in 
Eq.~(\ref{transosp1}) is not zero any more for $\vert D_{N,j}\rangle,$ while the denominator is zero and hence the state is not detected as entangled.
\bibitem{spin12} Here we assume that the state of a spin-$j$ 
particle can be mapped to a symmetric state of $2j$
spin-$\frac{1}{2}$ particles. In the appendix, there is a more detailed explanation.
\bibitem{sum} Note that here we symmetrized the correlation matrix. One can also define the antisymmetric part as $C^{\rm (A)}_{kl}:=\dfrac 1 2 \aver{J_k J_l-J_l J_k}.$ However, since for the commutator we have $[J_k,J_l]=iJ_m$ for $(k,l,m)=(x,y,z),$ $(y,z,x),$ and $(z,x,y),$ the antisymmetric part does not contain new information.
\end{thebibliography}
\end{document}